\PassOptionsToPackage{hypertexnames=false}{hyperref}
\documentclass[pdflatex,sn-nature]{sn-jnl}

\usepackage{graphicx}%
\usepackage{multirow}%
\usepackage{amsmath,amssymb,amsfonts}%
\usepackage{amsthm}%
\usepackage{mathrsfs}%
\usepackage[title]{appendix}%
\usepackage{xcolor}%
\usepackage{textcomp}%
\usepackage{manyfoot}%
\usepackage{booktabs}%
\usepackage{algorithm}%
\usepackage{algorithmicx}%
\usepackage{algpseudocode}%
\usepackage{listings}%

\usepackage{setspace}
\usepackage{natbib}
\usepackage{threeparttable}

\theoremstyle{thmstyleone}%
\theoremstyle{thmstyletwo}%
\theoremstyle{thmstylethree}%

\begin{document}

\title[Networked Risk Perception and Behavioral Bubbles]{Networked risk perception and behavioral bubbles: the case of a pandemic }

\author[1]{\fnm{Sepehr} \sur{Ilami}}

\author[2,3]{\fnm{Margherita} \sur{Comola}}

\author[1]{\fnm{Silvia} \sur{Prina}}

\author*[1]{\fnm{Babak} \sur{Heydari}}\email{b.heydari@northeastern.edu}

\affil*[1]{\orgname{Northeastern University}, \orgaddress{\city{Boston}, \state{MA}, \country{USA}}}

\affil[2]{\orgname{University of Paris-Saclay}, \orgaddress{\city{Paris}, \country{France}}}

\affil[3]{\orgname{Paris School of Economics}, \orgaddress{\city{Paris}, \country{France}}}

%%============================================================%%

\abstract{
Risk perception is typically modeled as an individual cognitive readout of objective hazard. Yet during crises what people judge as risky is shaped by what their peers do, making risk perception in part a property of communities, not just individuals. Using weekly mobility data from 313 Massachusetts municipalities over the first year of the COVID-19 pandemic and a pre-pandemic inter-town mobility network that fixes interaction structure before the shock, we estimate two-way fixed-effects panel regressions that separate the response to local case incidence, inter-town behavioral spillover along the mobility network, and within-town behavioral inertia; the pre-shock network and a lagged peer signal address the standard reflection and endogenous-group concerns. Inter-town behavioral spillovers are substantial and localize almost entirely within mobility-defined communities (the empirical referent of \emph{behavioral bubbles}), with effectively no propagation across community boundaries. The within-community spillover carries behavioral content beyond what peer-town case information alone would predict: when network-exposure-to-cases and network-exposure-to-behavior are raced, the behavioral channel survives and the case-exposure channel goes null. A four-way decomposition by mobility and demographic similarity shows the spillover requires both, concentrating where towns are routinely connected and socioeconomically similar and vanishing between similar towns that are not connected, which points to an observational and normative channel rather than a purely informational one. These results recast risk perception as a networked phenomenon and identify mobility-defined communities, rather than administrative units, as the operative scale of behavioral response. The pattern should generalize wherever exposure is uncertain, evolving, and socially negotiated, including climate adaptation and financial contagion.
}


\maketitle

\onehalfspacing

\section*{Introduction}
Low-probability, high-impact risks such as pandemics, climate shocks, and financial crises pose a persistent puzzle for science and policy \cite{kunreuther2013risk,aven2010risk}. Their objective properties can increasingly be quantified, yet societies routinely under-react to some threats and over-react to others \cite{KahnemanTversky1979,slovic2016perception,tversky1974judgment,slovic1987perception}. Even communities facing the same pathogen or physical hazard differ widely in how far they change daily routines, whether they trust institutions, and how much they invest in protection. During the COVID-19 pandemic, places with comparable age structures, health systems, and policy environments followed sharply different trajectories of behavior and mortality \cite{flaxman2020estimating,hsiang2020effect,haug2020ranking,brauner2021inferring,aboukheydari2021immediate}. Comparative work finds that even after demographics, comorbidities, socioeconomic composition, and the timing and stringency of interventions are accounted for, much of the cross-sectional variation in outcomes is left unexplained. That residual points to something systematic in how communities perceive and respond to risk that current models do not capture.

We argue that the missing structure is the social construction of \emph{risk perception} at the community level. Work in psychology, sociology, and behavioral economics has long shown that perceived risk is not a direct readout of objective probabilities but the product of cognitive heuristics, cultural frames, and social influence \cite{slovic1988risk,douglas1982can,KahnemanTversky1979,finucane2000affect,loewenstein2001risk,flynn1994gender}. The social amplification of risk framework describes how media, institutions, and interpersonal communication amplify or dampen hazard signals and so set off cascades of concern or complacency \cite{kasperson1988social,pidgeon2003social}. A parallel literature on social norms shows that people read their peers to judge what is safe, appropriate, or necessary, and adjust accordingly \cite{cialdini2004social,bicchieri2005grammar,fehr2004social,cialdini1990focus}. Perceived risk, on this view, is negotiated and reinforced within communities, and it can stay sticky or path-dependent even as objective conditions move \cite{brewer2007meta,savadori2021risk,lavell2020social,tsoy2021role}. Yet most formal models of behavior under risk, whether in economics, epidemiology, or policy analysis, still treat risk perception as an individual attribute \cite{bauch2013social,reluga2010game,funk2010modelling,fenichel2011adaptive}. They picture individuals weighing costs against local risk, sometimes with bounded rationality or prospect-theoretic distortions, and leave the social processes that generate perceived risk implicit. This abstracts away a regularity that crises make hard to ignore: much of what people do is coordinated and constrained at the group level \cite{centola2010spread,centola2018behavior,caoheydari2022micro}.

Here we give direct empirical evidence that community-level peer effects matter for behavior under risk. People do not meet pandemic, climate, or financial risk alone. They meet it as members of communities whose members watch one another, trade information, and settle on implicit thresholds of how much caution is enough \cite{bavel2020using,betsch2020behavioural}. Using high-frequency mobility data from 313 Massachusetts municipalities across the first pandemic year, we show that a town's behavior responds to its local infection rate and, to a substantial degree, to what its mobility-linked peers are doing \cite{kraemer2020effect,chang2021mobility,chan2020risk,barbieri2021impact,lawal2022movement,galasso2020gender}. We build a pre-pandemic mobility network from device-level travel flows recorded before the state's first COVID-19 case and treat it as an exogenous structure along which behavioral norms can travel \cite{colizza2007reaction,balcan2009multiscale,brockmann2013hidden,meloni2011modeling,pastor2015epidemic}. The design separates three forces acting on town mobility. The first is local objective risk, the response of a town's behavior to its own weekly incidence. The second is within-town inertia, which at the town level folds residents' own-behavior persistence together with within-town peer reinforcement into one lagged-outcome channel. The third is across-town spillover, carried by pre-existing interaction patterns and, as we show, gated by demographic similarity.

Three findings anchor the paper. The first is that once we condition on local case incidence, time-invariant town characteristics, and common weekly shocks, the cross-community variation that remains in protective behavior is itself organized along the pre-pandemic mobility network. The second is that this structure is local: a Louvain partition of the network \cite{blondel2008fast} splits the spillover into a within-community component that carries the signal and a across-community component statistically indistinguishable from zero, the empirical referent of what we call behavioral bubbles. The third characterizes when the spillover operates. Crossing the mobility partition with a near-orthogonal partition on demographic similarity shows that the channel requires both. It is concentrated among towns that are routinely connected and demographically similar, it is absent across mobility-community boundaries even when the towns on either side resemble one another, and it is only partial among connected towns that differ demographically. Two things follow. Routine connection is necessary but not sufficient, so demographic similarity gates the network channel rather than acting as a separate driver. And the pattern rules out the most natural alternative to social influence, that demographically similar places simply react alike to similar objective conditions: were that the case, similar but unconnected towns would co-move, and they do not. Two further results round out the picture. Towns respond to local risk in the direction any precautionary model predicts. And behavior is persistent at the town level, so the lagged-outcome channel, working together with the peer channel, hardens early differences into lasting divergence; a community that settles into caution, or into normality, tends to stay there absent a substantial local or network shock \cite{manski1993identification,bramoulle2009identification,jackson2008social}.

We read these results as micro-level evidence that risk perception is socially constructed at the community level. Rather than respond only to its own epidemiological situation, a community appears to infer how much caution is warranted from its interaction network, aligning in part with what its mobility-linked and demographically similar peers do. This helps explain why so much cross-place variation in pandemic outcomes resists demographics, institutions, and formal policy \cite{flaxman2020estimating,hsiang2020effect}. It also suggests that models of collective response to systemic risk should treat perceived risk as a property of interacting communities rather than of isolated individuals \cite{bavel2020using,betsch2020behavioural}, and that interventions built around administrative units, or on an assumption of independent response, can miss the channels through which perception and behavior actually move.

COVID-19 is our setting, but the mechanism is not specific to it. Behavioral spillovers along social and spatial networks should arise whenever risks are uncertain, evolving, and socially contested, conditions that describe climate adaptation, financial contagion, and technological hazards as much as infectious disease. The pandemic simply supplies an unusually rich, high-frequency record of both behavior and exposure. What it reveals is portable, and the sections that follow make the case that perceived risk has a measurable community-level structure that current models leave out.

\begin{figure}[!htbp]
    \centering
    \includegraphics[width=\linewidth]{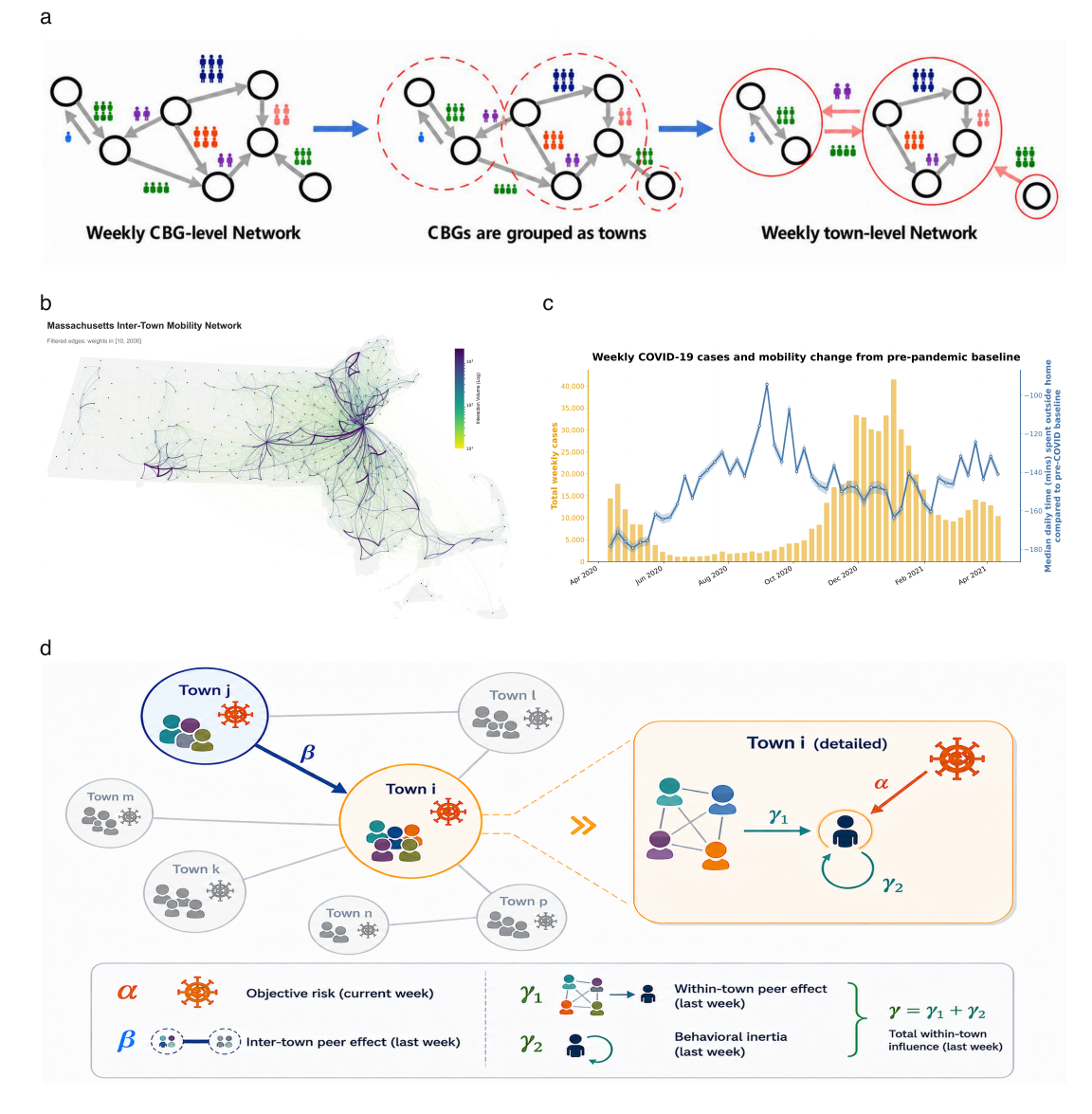}
    \caption{\textbf{Data, setting, and conceptual framework.}
    \textbf{(a)} Construction of the town-level mobility network: SafeGraph device flows between census block groups (CBGs) are aggregated to census tracts and then to towns, yielding a town-level mobility network that captures routine interaction patterns.
    \textbf{(b)} The resulting Massachusetts inter-town mobility network ($N=313$ towns), with edge thickness and color encoding flow volume on a log scale.
    \textbf{(c)} State-level time series over the first pandemic year: weekly confirmed COVID-19 cases (bars, left axis) against average non-home dwell time (line, right axis), motivating a decomposition of behavior that \textit{goes beyond local case counts}.
    \textbf{(d)} Conceptual model of the three forces shaping a focal town's mobility, with the regression coefficients identified for each: local objective risk ($\alpha$, weekly cases per $10{,}000$, $X_{i,t}$), inter-town behavioral spillover along the pre-pandemic mobility network ($\beta$, on $S^{\mathrm{PreG}}_{i,t-1} = \sum_j \mathrm{PreG}_{ij}\,Y^{adj}_{j,t-1}$; decomposed in Results into within- and across-mobility-community pairs), and within-town behavioral inertia ($\gamma$, on the lagged outcome $Y^{adj}_{i,t-1}$). Because the unit of analysis is the town, $\gamma$ does not isolate purely individual-level persistence: at this aggregation it combines residents' own-behavior persistence with within-town peer reinforcement, both of which propagate from $t{-}1$ to $t$ through the same lagged-outcome channel. The inter-town peer effects (across distinct towns) are captured separately by $\beta$ and its decompositions. Town and week fixed effects, denoted $\mu_i$ and $\lambda_t$ in the regression equation (Eq.~\ref{eq:headline}), are not depicted.}
    \label{fig:ensemble_figure_1}
\end{figure}

\section*{Results}

\subsection*{Setting, outcome, and the pre-pandemic mobility network}

We test the social-construction hypothesis on town-level mobility data from Massachusetts during the first year of the COVID-19 pandemic. The setting offers cross-sectional variation across 313 municipalities while holding the state-level legal and institutional environment fixed, and weekly device-flow data make it possible to track behavior at a frequency comparable to the timescale at which information and norms diffuse. The outcome $Y^{adj}_{i,t}$ is the median daily non-home dwell time in town $i$ during week $t$, expressed as a deviation from the town's last two pre-pandemic weeks: positive values mean residents spent longer outside the home than they did pre-COVID, negative values the reverse. Local hazard is captured by the weekly case rate per 10{,}000 residents, $X_{i,t}$.

The empirical strategy hinges on a pre-determined network along which behavioral signals can propagate. We construct $\mathrm{PreG}$ from the same device-flow data restricted to the last two weeks of January 2020 (before the first confirmed case in the state) and normalize each row by its outflow total, so that $\mathrm{PreG}_{ij}$ is the share of town $i$'s routine outflow directed toward town $j$. Because the network predates any pandemic exposure, it captures the state's standing travel structure of commuting corridors, retail catchments, and social ties without itself being shaped by the pandemic, and it serves as a pre-determined channel along which behavioral signals can cross between towns. The principal spillover regressor is
\begin{equation}
S^{\mathrm{PreG}}_{i,t-1} \;=\; \sum_{j} \mathrm{PreG}_{ij}\, Y^{adj}_{j,t-1},
\label{eq:s_preg}
\end{equation}
the network-weighted lagged behavior of town $i$'s routine peers. Throughout we estimate two-way fixed-effects panel regressions with town fixed effects (absorbing time-invariant municipal characteristics) and week fixed effects (absorbing common shocks such as state-wide policy and national news), with standard errors clustered at the town level. Methodological detail is deferred to Methods.

\subsection*{Peer behavior structures the cross-community response}

\begin{figure}[!htbp]
    \centering
    \includegraphics[width=\textwidth]{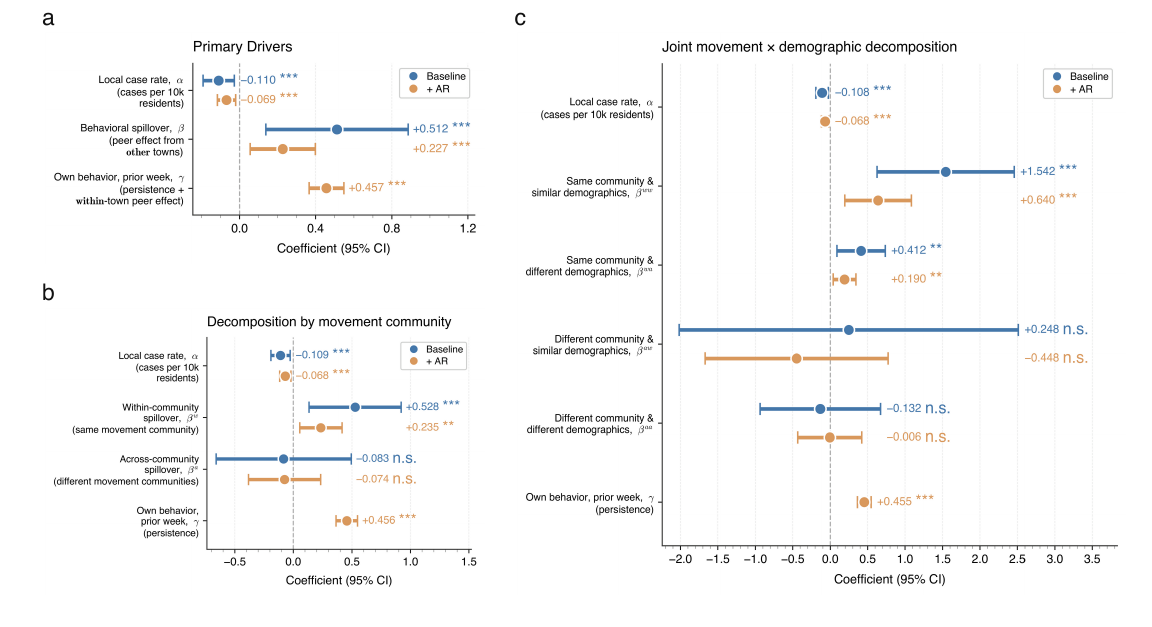}
    \caption{\textbf{Inter-town peer effect, decomposed by community structure: baseline vs. AR(1).} Each panel pairs the FE-only baseline (blue) with the FE\,+\,AR(1) specification (orange); markers are point estimates, bars are $95\%$ confidence intervals. \textbf{a}, Headline three-driver decomposition (Eq.~\eqref{eq:headline}): the local hazard response $\alpha$ on $X_{i,t}$, the inter-town peer effect $\beta$ on $S^{\mathrm{PreG}}_{i,t-1}$, and the own-lag coefficient $\gamma$ on $Y^{adj}_{i,t-1}$ (Table~\ref{tab:results_main}, cols.~1--2). $\gamma$ enters only in the AR(1) specification, so its baseline marker is omitted. \textbf{b}, Mobility-community decomposition of the peer-effect channel under the mobility-Louvain partition (Fig.~\ref{fig:partitions}a): the within-community peer effect $\beta^{w}$ vs.\ the across-community peer effect $\beta^{a}$ (Table~\ref{tab:results_main}, cols.~3--4). The within-community channel carries the spillover signal; the across-community channel is statistically indistinguishable from zero. \textbf{c}, Joint mobility\,$\times$\,demographic decomposition: $\beta^{ww}$ (same mobility community, similar demographics), $\beta^{wa}$ (same mobility, different demographics), $\beta^{aw}$ (different mobility, similar demographics), and $\beta^{aa}$ (different on both). Peer effects concentrate in the within-mobility-community cells, and within those, in the demographically similar partition. All models include town and week fixed effects; standard errors are clustered at the town level. Significance: $^{*}p<0.10$, $^{**}p<0.05$, $^{***}p<0.01$; ``n.s.'' otherwise.}
    \label{fig:results_main}
\end{figure}

Table~\ref{tab:results_main} (columns~1--2) and Fig.~\ref{fig:results_main}\textbf{a},\textbf{b} report the three structural coefficients of Eq.~\eqref{eq:headline}: $\alpha$ on local hazard $X_{i,t}$, $\beta$ on the inter-town peer signal $S^{\mathrm{PreG}}_{i,t-1}$, and $\gamma$ on the own lag $Y^{adj}_{i,t-1}$. Each coefficient is significant in the headline specification, and each plays a distinct role.

Towns do respond to local risk in the direction any precautionary model predicts: $\hat\alpha < 0$, with higher case rates predicting reduced non-home dwell time. At the winter 2020--21 peaks, when weekly incidence in the worst-hit towns ran above $100$ cases per $10{,}000$, the implied hazard-driven reduction in non-home dwell time is on the order of a few minutes per day relative to a town with near-zero incidence.

The inter-town peer signal $\beta$ carries the main result, and its magnitude depends on whether own-town persistence is controlled, so the two pooled columns must be read together. Without the own lag (column~1) $\hat\beta = 0.51$; with it (column~2) $\hat\beta = 0.23$ ($p<0.01$). The drop is not the peer effect weakening under scrutiny but the removal of a known upward bias: the omitted own-town behavior $Y^{adj}_{i,t-1}$ is positively correlated with the peer signal, because connected towns co-move, so the static estimate absorbs part of a town's own persistence and overstates the across-town channel. The FE+AR(1) estimate $\hat\beta = 0.23$ is the clean across-town response, the share of a one-unit shift in the network-weighted lagged behavior of a town's routine peers that passes into its own behavior the following week, holding own-town momentum, contemporaneous hazard, and time-invariant town characteristics fixed. This partialed estimate is the one that carries the interpretation, and it is a lower-bound measure of social influence: it captures only the across-town channel, because any peer reinforcement within the focal town has nowhere to go and is absorbed into the lagged-outcome term $\gamma\,Y^{adj}_{i,t-1}$, since the town aggregation cannot separate residents' own persistence from within-town peer adjustment \cite{manski1993identification,bramoulle2009identification,jackson2008social}.

The within-town channel $\gamma$ supplies the matching upper bound. With $\hat\gamma = 0.46$, the lagged-outcome term folds individual persistence together with within-town peer reinforcement at the town level, so the sum $\hat\beta + \hat\gamma = 0.68$ is an upper-bound measure of behavior-on-behavior coupling per unit lagged behavior, taking in every within-town and across-town peer channel along with the residual individual persistence that cannot be separated out without individual-level ties. The true social-influence response per unit peer behavior therefore lies between $\hat\beta = 0.23$ (across-town only) and $\hat\beta + \hat\gamma = 0.68$. These are one-week responses. Because the own lag re-injects part of last week's peer-driven shift, the cumulative own-town response to a sustained change in peer behavior is the dynamic multiplier $\hat\beta / (1 - \hat\gamma) \approx 0.42$, larger than the one-step $\hat\beta$ but still below the inflated static estimate of $0.51$, which therefore overstates even the long-run peer effect and is not itself a bound.

Taken together, the headline specification isolates a substantial network-mediated channel in the cross-community variation that local hazard and town characteristics leave unexplained. We take up what this residual structure implies for behavioral science, for adaptive epidemic models, and for intervention design in the Discussion.

Behavior is also persistent at the town level, and the persistence carries content. The within-town channel $\gamma$ implies an own-town-shock half-life of about a week on its own, and the inter-town channel lengthens that decay because lagged peer behavior re-injects last week's shock at each step. The effective half-life of a community-level shock, governed by the dominant eigenvalue of $\hat\gamma\,I + \hat\beta\,\mathrm{PreG}$ acting on the within-mobility-community block, runs materially longer than the AR(1) parameter alone would suggest. Two towns with nearly identical objective exposure but different early shocks then settle onto persistently divergent trajectories. Once a community converges on caution or on normality, the joint action of within-town inertia and the within-community peer effect $\beta^{w}$ holds it there.

\begin{table}[!htbp]
\centering
\caption{\textbf{Inter-town behavioral spillover along the pre-pandemic mobility network, pooled and decomposed.} Outcome $Y^{adj}_{i,t}$: weekly median non-home dwell time minus the town's pre-pandemic baseline (minutes per day). $X_{i,t}$: weekly cases per $10{,}000$. $S^{\mathrm{PreG}}_{i,t-1}$: network-weighted lagged behavior of pre-pandemic mobility peers (Eq.~\ref{eq:s_preg}). $S^{\mathrm{PreG},w}$ and $S^{\mathrm{PreG},b}$: Hadamard-masked versions under the mobility-Louvain partition (Fig.~\ref{fig:partitions}a). $S^{\mathrm{Demo},w}$ and $S^{\mathrm{Demo},b}$: analogous masks under the demographic-Louvain partition (Fig.~\ref{fig:partitions}c). $S^{\mathrm{PreG},ww}$, $S^{\mathrm{PreG},wa}$, $S^{\mathrm{PreG},aw}$, $S^{\mathrm{PreG},aa}$: the four masks crossing the two partitions, indexed by (within/across mobility, within/across demographic), matching the four-way decomposition in Fig.~\ref{fig:results_main}c. All columns include town and week fixed effects; standard errors clustered at the town level in parentheses. $^{*}p<0.10$, $^{**}p<0.05$, $^{***}p<0.01$.}
\label{tab:results_main}
\fontsize{7}{8.2}\selectfont
\setlength{\tabcolsep}{0pt}
\begin{tabular*}{\linewidth}{@{\extracolsep{\fill}} l cc | cc | cc | cc}
\toprule
 & \multicolumn{8}{c}{\textit{Dep.\ var.: $Y^{adj}_{i,t}$ (min/day)}} \\
\cmidrule(lr){2-9}
 & \multicolumn{2}{c|}{Pooled} & \multicolumn{2}{c|}{Mobility w/a} & \multicolumn{2}{c|}{Demographic w/a} & \multicolumn{2}{c}{Mobility $\times$ Demo} \\
\cmidrule(lr){2-3} \cmidrule(lr){4-5} \cmidrule(lr){6-7} \cmidrule(lr){8-9}
 & (1) & (2) & (3) & (4) & (5) & (6) & (7) & (8) \\
 & FE & +AR(1) & FE & +AR(1) & FE & +AR(1) & FE & +AR(1) \\
\midrule
$Y^{adj}_{i,t-1}$ & & $0.457^{***}$ & & $0.456^{***}$ & & $0.455^{***}$ & & $0.455^{***}$ \\
 & & $(0.046)$ & & $(0.046)$ & & $(0.047)$ & & $(0.047)$ \\[3pt]
$X_{i,t}$ & $-0.110^{***}$ & $-0.069^{***}$ & $-0.109^{***}$ & $-0.068^{***}$ & $-0.109^{***}$ & $-0.069^{***}$ & $-0.108^{***}$ & $-0.068^{***}$ \\
 & $(0.042)$ & $(0.025)$ & $(0.042)$ & $(0.025)$ & $(0.042)$ & $(0.024)$ & $(0.042)$ & $(0.025)$ \\[3pt]
$S^{\mathrm{PreG}}_{i,t-1}$ & $0.512^{***}$ & $0.227^{***}$ & & & & & & \\
 & $(0.191)$ & $(0.088)$ & & & & & & \\[3pt]
$S^{\mathrm{PreG},w}_{i,t-1}$ & & & $0.528^{***}$ & $0.235^{**}$ & & & & \\
 & & & $(0.200)$ & $(0.092)$ & & & & \\[3pt]
$S^{\mathrm{PreG},a}_{i,t-1}$ & & & $-0.083$ & $-0.074$ & & & & \\
 & & & $(0.294)$ & $(0.157)$ & & & & \\[3pt]
$S^{\mathrm{Demo},w}_{i,t-1}$ & & & & & $1.499^{***}$ & $0.608^{***}$ & & \\
 & & & & & $(0.462)$ & $(0.222)$ & & \\[3pt]
$S^{\mathrm{Demo},a}_{i,t-1}$ & & & & & $0.399^{**}$ & $0.185^{**}$ & & \\
 & & & & & $(0.156)$ & $(0.075)$ & & \\[3pt]
$S^{\mathrm{PreG},ww}_{i,t-1}$ & & & & & & & $1.542^{***}$ & $0.640^{***}$ \\
 & & & & & & & $(0.468)$ & $(0.228)$ \\[3pt]
$S^{\mathrm{PreG},wa}_{i,t-1}$ & & & & & & & $0.412^{**}$ & $0.190^{**}$ \\
 & & & & & & & $(0.165)$ & $(0.078)$ \\[3pt]
$S^{\mathrm{PreG},aw}_{i,t-1}$ & & & & & & & $0.248$ & $-0.448$ \\
 & & & & & & & $(1.156)$ & $(0.623)$ \\[3pt]
$S^{\mathrm{PreG},aa}_{i,t-1}$ & & & & & & & $-0.132$ & $-0.006$ \\
 & & & & & & & $(0.410)$ & $(0.218)$ \\
\midrule
$R^{2}$            & 0.772 & 0.822 & 0.772 & 0.822 & 0.773 & 0.822 & 0.773 & 0.822 \\
$N$ (town-weeks)   & 16{,}276 & 16{,}276 & 16{,}276 & 16{,}276 & 16{,}276 & 16{,}276 & 16{,}276 & 16{,}276 \\
Town FE            & Yes & Yes & Yes & Yes & Yes & Yes & Yes & Yes \\
Week FE            & Yes & Yes & Yes & Yes & Yes & Yes & Yes & Yes \\
\bottomrule
\end{tabular*}
\vspace{4pt}
{\scriptsize \textit{Notes.} Columns 1--2 report the pooled network spillover with and without the lagged outcome. Columns 3--4 split it into within- and across-community pairs of the mobility-Louvain partition: the within term carries the channel and the across term is statistically indistinguishable from zero, both with and without persistence. Columns 5--6 report the analogous split under the demographic-Louvain partition: both within and across components enter positively, indicating that the demographic partition does not, on its own, carve up the network channel. Columns 7--8 report the joint four-way decomposition crossing the mobility and demographic partitions (Fig.~\ref{fig:results_main}c): the spillover concentrates in the within-mobility cells---strongest where towns also share a demographic community ($S^{\mathrm{PreG},ww}$) and weaker but still significant across demographics ($S^{\mathrm{PreG},wa}$)---while both across-mobility terms are statistically indistinguishable from zero. SI Tables~\ref{tab:si_mobility_partition} and~\ref{tab:si_demographic_partition} report auxiliary specifications that re-introduce the unsplit $S^{\mathrm{PreG}}_{i,t-1}$ alongside the masked components.}
\end{table}

\subsection*{Inter-town spillover localizes within mobility communities}

\begin{figure}[!htbp]
    \centering
    \includegraphics[width=\textwidth]{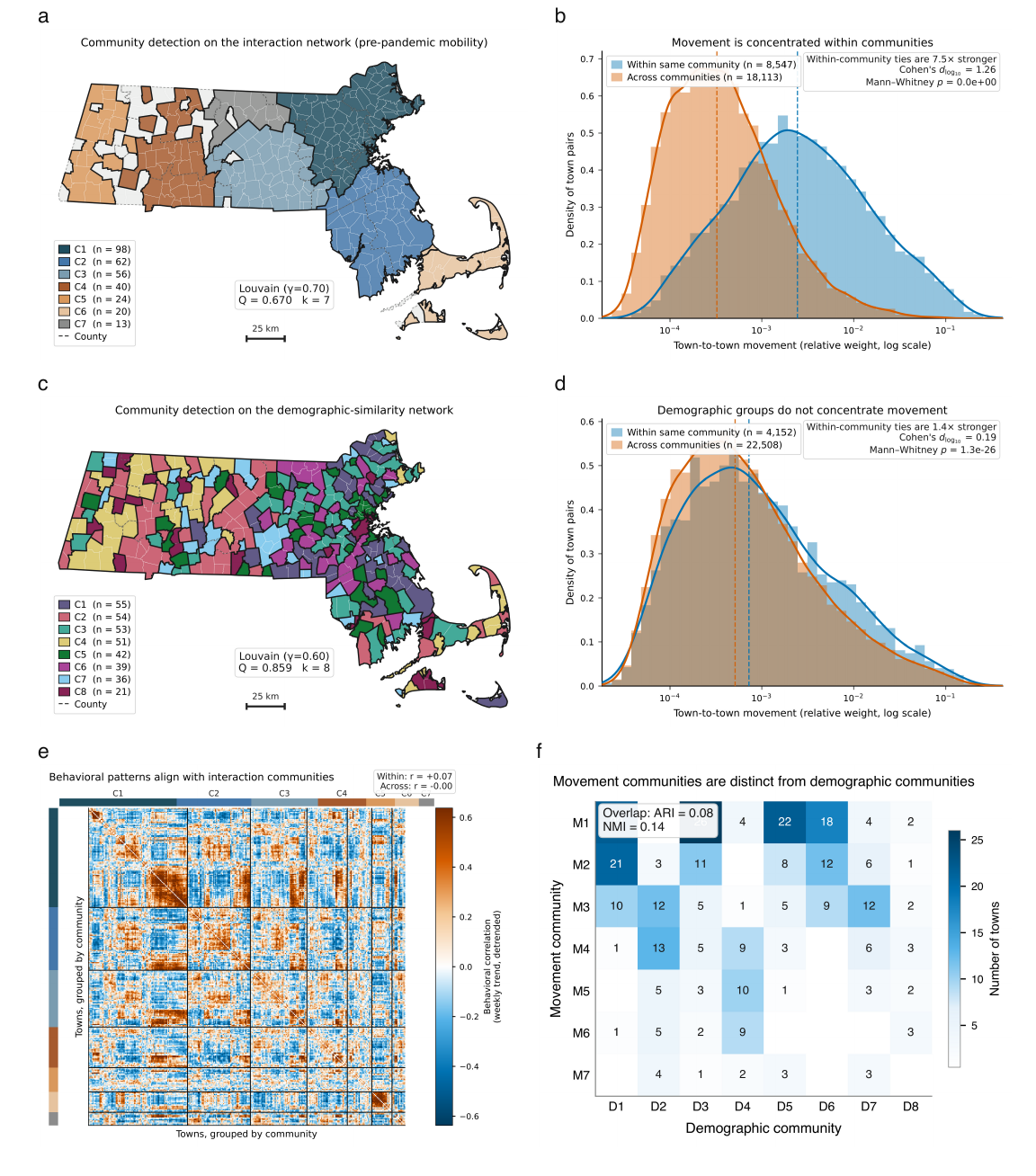}
    \caption{\textbf{Mobility and demographic community structure of Massachusetts towns.}
    \textbf{a}, Mobility-Louvain partition of the pre-pandemic flow matrix $\mathrm{PreG}$ ($k = 7$, $Q = 0.67$); thick outlines mark community boundaries, dashed grey lines mark county boundaries.
    \textbf{b}, Within- versus across-community $\log_{10}\mathrm{PreG}$ edge-weight distribution under the mobility partition: within-mobility edges are systematically heavier.
    \textbf{c}, Demographic-Louvain partition built on four town-level indicators (median age, net worth, diversity index, population density), $k = 8$, $Q \approx 0.86$; used as the demographic axis of the four-way decomposition in Fig.~\ref{fig:results_main}\textbf{c}.
    \textbf{d}, Within- versus across-community $\mathrm{PreG}$ distribution under the demographic partition; in contrast to \textbf{b}, demographic similarity does not, on its own, imply heavier routine flow.
    \textbf{e}, Pairwise Pearson correlation of week-detrended $Y^{\mathrm{adj}}$ trajectories, ordered by mobility-Louvain community. The block-diagonal structure shows the mobility partition organizes behavioral co-movement, not only connectivity.
    \textbf{f}, Mobility\,$\times$\,demographic Louvain contingency on the 313-town intersection ($\mathrm{ARI} = 0.08$, $\mathrm{NMI} = 0.14$). The two partitions are essentially orthogonal, which is what makes the four-way decomposition non-redundant.
    Methodology, resolution-sweep tuning, and a Clauset--Newman--Moore greedy-modularity baseline are detailed in Methods §\ref{sec:methods-communities} and SI §\ref{suppsubsec:greedy-cd}.}
    \label{fig:partitions}
\end{figure}

Where does the inter-town peer effect travel? A Louvain modularity partition of the symmetrized origin-normalized $\mathrm{PreG}$ recovers seven mobility communities (Fig.~\ref{fig:partitions}\textbf{a}). They are spatially coherent, picking out a Greater Boston cluster, a Cape Cod and South Coast cluster, a Pioneer Valley cluster, a Berkshires cluster, and so on, yet they do not track the Massachusetts county boundaries overlaid as dashed lines. The partition is organized by routine flow, not by jurisdictional design.

The mobility partition matters for behavior as well as for connectivity. Within-community $\mathrm{PreG}$ edges are markedly heavier than across-community edges (Fig.~\ref{fig:partitions}\textbf{b}), and the same partition organizes behavioral co-movement, not connectivity alone. After we remove the common pandemic shock by week-demeaning each town's $Y^{adj}$ trajectory, the pairwise Pearson correlations across towns, ordered by community membership, fall into a clear block-diagonal pattern (Fig.~\ref{fig:partitions}\textbf{e}): positive within mobility communities and essentially zero across them. A partition recovered from pre-pandemic flows alone, with no reference to outcomes, thus accounts for a meaningful share of behavioral co-movement during the pandemic.

A complementary Louvain partition on four widely used town-level demographic indicators, namely median age, net worth, diversity index, and population density, recovers eight demographic communities (Fig.~\ref{fig:partitions}\textbf{c}). Two properties matter for what follows. First, these communities are not connectivity-coherent: within-demographic-community pairs are no heavier in $\mathrm{PreG}$ than across-community pairs (Fig.~\ref{fig:partitions}\textbf{d}), so demographically similar towns are not the towns that travel between one another. Second, on the 313-town intersection the mobility and demographic partitions are close to orthogonal ($\mathrm{ARI} = 0.08$, $\mathrm{NMI} = 0.14$; Fig.~\ref{fig:partitions}\textbf{f}), so membership in a demographic community is not redundant with membership in a mobility community. Crossing the two therefore produces four genuinely distinct kinds of inter-town pair rather than a relabeling of the mobility split, which is the structural basis for the four-way decomposition below.

The mobility partition cleanly splits the inter-town peer effect. Masking $\mathrm{PreG}$ element-wise with the within- and across-community indicators (Methods §\ref{sec:methods-masks}) gives two spillover regressors that enter as separate terms (Table~\ref{tab:results_main}, columns~3--4; Fig.~\ref{fig:results_main}\textbf{b}). The within-community coefficient $\beta^{w}$ is large and significant; the across-community coefficient $\beta^{a}$ is statistically indistinguishable from zero. The split holds once within-town inertia is absorbed in column~4, and auxiliary specifications that re-introduce the unsplit $S^{\mathrm{PreG}}_{i,t-1}$ alongside the masked components confirm that the within-community pairs carry the channel (SI Table~\ref{tab:si_mobility_partition}). The peer effect, already a lower-bound measure of social influence, runs almost entirely along edges inside the same routine-interaction community. This is the empirical referent of a behavioral bubble \cite{centola2010spread,centola2018behavior}.

\subsection*{Four-way mobility-by-demographic decomposition}

Crossing the mobility and demographic Louvain partitions splits $\mathrm{PreG}$ into four Hadamard-masked components, indexed by whether each town pair is within the same or across different mobility communities and within the same or across different demographic communities (Methods §\ref{sec:methods-masks}). The four spillover regressors enter Eq.~\eqref{eq:headline} as separate terms, with coefficients reported in Fig.~\ref{fig:results_main}\textbf{c}. The peer effect concentrates in one block: $\beta^{ww}$, on pairs sharing both a mobility community and a demographic community, is several times the pooled $\hat\beta$; $\beta^{wa}$, on pairs in the same mobility community but different demographic communities, is smaller, though still positive and significant; and the two across-mobility coefficients, $\beta^{aw}$ and $\beta^{aa}$, are statistically indistinguishable from zero. The spillover thus needs both kinds of proximity. Routine connection is necessary, since across mobility-community boundaries we detect no spillover whether or not the two towns are demographically similar. Connection on its own is not sufficient, since among connected towns the channel concentrates in the demographically similar pairs. Demographic similarity gates the mobility channel rather than carrying one of its own.

The null across-mobility coefficients also bear on identification. The most natural alternative to social influence is that demographically similar towns respond in parallel to similar objective conditions, a pattern that could masquerade as network spillover. If that were the source, demographically similar but unconnected towns would co-move; that is the $\beta^{aw}$ block, and it is indistinguishable from zero. The crossing is what isolates this test. A decomposition along demographic similarity alone loads on both its within and across blocks (Table~\ref{tab:results_main}, columns~5--6), because that partition cuts across mobility communities (Fig.~\ref{fig:partitions}\textbf{f}), so only the joint split separates out the similar-but-unconnected pairs the confound would require. A continuous-modulator test that replaces the binary demographic mask with the town-by-town similarity matrix $D$ and races $\mathrm{PreG}\odot D$ against $\mathrm{PreG}$ (SI §\ref{suppsubsec:methods-D}, SI Table~\ref{tab:results_demographic}) reaches the same conclusion.

\subsection*{Robustness and falsifications}

Three checks address the alternative readings most likely to worry a reader. First, the pattern is not an artifact of a few large municipalities. Dropping the 32 towns above the $90^{\text{th}}$ population percentile, among them Boston, Worcester, and Springfield, leaves $281$ towns and $14{,}612$ town-weeks, and in that reduced sample the spillover signature holds in magnitude and significance across every specification in Table~\ref{tab:results_main} (SI Table~\ref{tab:models_with_excluded_towns}). It is not driven by metropolitan heterogeneity or by Boston's unusual position in the network.

Second, the spillover is a behavioral channel rather than a cases-aware one. Replacing $S^{\mathrm{PreG}}_{i,t-1}$ with a network-exposure-to-cases term $S^{\mathrm{PreG}\cdot X}_{i,t-1} = \sum_j \mathrm{PreG}_{ij}\,X_{j,t-1}$, or entering the two together, leaves the cases term statistically indistinguishable from zero while $S^{\mathrm{PreG}}_{i,t-1}$ keeps its sign and magnitude (SI Table~\ref{tab:models2_with_excluded_towns}). Towns adjust to peer behavior, not to peer hazard, which fits norm-based and observational-learning accounts of diffusion better than a purely informational risk-news channel.

Third, the spillover is specific to the mobility network rather than a generic spatial-correlation artifact. Re-estimating it on a geographic-proximity placebo, an inverse-distance or shared-county adjacency that encodes proximity without routine interaction, in place of $\mathrm{PreG}$ separates the mobility-specific content of the channel from spatial autocorrelation; a channel that works through routine interaction predicts an attenuated placebo coefficient relative to the mobility-network estimate.

\section*{Discussion}

Town-level protective behavior in the first pandemic year reflects two things: a response to local case incidence, and a residual cross-community variation that local hazard and time-invariant town characteristics do not explain. That residual is the object standard models assign to individual risk perception, and our central result is that it is itself structured. It tracks peer behavior along the pre-pandemic mobility network, and it does so within mobility communities in particular: the spillover runs along edges that connect towns inside the same routine-interaction cluster and is statistically absent across community boundaries, the signature we call a behavioral bubble. The across-town coefficient $\beta$ is a lower-bound measure of social influence, since within-town peer reinforcement is absorbed into the persistence term $\gamma$, and the matching upper bound $\beta + \gamma$ points the same way. The demographic decomposition in the SI (SI §\ref{suppsec:demographic-robustness}) sharpens the pattern rather than weakening it, strongest where peers are demographically as well as routinely similar.

Why does the residual take this particular shape, organized along peer behavior rather than along peer hazard or noise? The candidate inputs to community-level risk calibration are not equally legible. Local incidence is abstract, lagged, and noisily measured: a weekly case rate per ten thousand residents depends on testing intensity, arrives with delay, and is hard for any resident to turn into a felt sense of personal exposure. Peer behavior is immediate and directly visible, in whether neighbors are out, whether the center of town is busy or shuttered, whether connected communities have loosened or tightened. When residents judge how much caution is enough, they appear to anchor on the legible, continuously updated cue. Two features of our results fit this account. The channel loads on peer behavior and not on peer cases: exposure to connected towns' incidence adds nothing once their behavior is in the model, which is what an observational or normative mechanism predicts and a purely informational one does not. And the signal concentrates where peers are actually observed, inside routine-interaction communities, rather than diffusing across the state. Read this way, the shape of community-level risk perception is a reasonable response to which cue is easier to see, and it carries a cost: it locks communities onto divergent paths that their objective circumstances alone would not produce.

The broader contribution is to move risk perception from the individual to the community. The social-amplification-of-risk \cite{kasperson1988social,pidgeon2003social} and social-norm \cite{cialdini2004social,bicchieri2005grammar} traditions long held that perceived risk is socially negotiated, and network science established that behavior spreads along social ties \cite{centola2010spread,centola2018behavior,centola2007complex,christakis2007spread,granovetter1978threshold,watts2002simple,bond2012experiment,aral2012identifying,centola2011experimental}. What has been missing is a way to measure that negotiation at the scale of geographic communities. Our estimates provide one: after conditioning on local hazard and town characteristics, the residual cross-community variation in behavior is organized along persistent routine-interaction channels, concentrated within mobility-defined communities and gated by demographic similarity. That the channel requires both a routine tie and demographic similarity, rather than connection alone, fits accounts in which social influence operates through perceived relevance: a community appears to weight the behavior of peers it both interacts with and resembles, reinforced among similar actors rather than transmitted by contact alone \cite{centola2007complex,centola2011experimental}. Risk perception, in this sense, has a measurable community-level shape, one that can be partitioned and put to use. The pre-determined network and lagged-peer design recover the across-town spillover where the reflection problem would otherwise block identification, and they bound the within-town channel rather than discarding it \cite{manski1993identification,bramoulle2009identification,sacerdote2001peer,bramoulle2020peer,comolaprina2021}.

The modeling implication is direct. Epidemic models already route pathogens along mobility networks \cite{colizza2007reaction,balcan2009multiscale,brockmann2013hidden,kraemer2020effect,chang2021mobility}; our results show that the same network also routes caution, coupling disease and behavioral dynamics on one substrate. A behaviorally adaptive model should therefore carry precaution as a network-mediated state, updating on a peer-weighted average of peers' precaution with weights set by mobility flow and gated by demographic similarity, rather than as a private function of local prevalence. This is the network structure that behavioral-feedback epidemic models have so far had to assume \cite{bauch2013social,reluga2010game,fenichel2011adaptive,funk2010modelling,verelst2016behavioural,bedford2019new,rahmandad2021behavioral,rahmandad2022missing}, and a model that leaves it out will misjudge the reach of communication-based interventions.

The same structure changes how interventions should be targeted. Because behavior co-moves within mobility communities that cut across county and health-district lines (Fig.~\ref{fig:partitions}\textbf{a}), communication organized around administrative units will miss the scale on which behavior actually moves; and because the structured part of behavior runs through peers, messaging aimed only at where cases are will reach less far than messaging aimed at where peers are. That points to a concrete lever: within a mobility community, moving a few influential towns can shift the whole cluster through the within-community channel $\beta^{w}$, whereas effort allocated on the case distribution alone is spent on towns whose behavior is being set by peers outside the targeting unit. It also gives a readable diagnostic. The share of cross-place outcome variation left unexplained by demographics, comorbidities, and formal policy \cite{flaxman2020estimating,hsiang2020effect,allcott2020polarization,goolsbee2021fear,painter2021political,holtz2020interdependence,devaan2021social,caoheydari2022micro} depends in part on where a place sits in the mobility-community structure, a quantity recoverable from device flows and census data before the next shock. Related work on the same Massachusetts towns finds that a town's mobility-network position carries out-of-sample predictive value for short-horizon case forecasts, largest where granular local case histories are coarse or delayed \cite{ilami2025network}.

Two limitations bind the causal reading most tightly. First, the design is ecological: spillovers are estimated over towns, not individuals, so the town-level result is consistent with several individual mechanisms, among them observational learning, normative conformity, and shared exposure to local media, which we cannot adjudicate among, although the behavior-not-hazard and within-community patterns both lean away from a purely informational account. Second, while the lagged-peer specification and the pre-pandemic network neutralize the most direct identification threats, namely Manski reflection and endogenous group formation, they cannot rule out a time-varying unobserved shock that travels along the same edges as $\mathrm{PreG}$; the network-exposure-to-cases falsification and the geographic-proximity placebo address this where a counter-structure can be built, and the four-way decomposition adds another: a shared-conditions account, in which demographically similar towns simply react alike to correlated objective risk, would produce co-movement between similar towns whether or not they are connected, yet the similar-but-unconnected block is null. A fully causal claim would still need a behavioral instrument we do not have. The remaining caveats are narrower. Non-home dwell time is one window on protective behavior and leaves masking and contact patterns unobserved; SafeGraph coverage underweights residents without smartphones and drops 38 small rural towns; and the geography of Massachusetts, one dominant metro beside a rural west, may not carry over to states with several comparable metros, even though the empirical design itself travels wherever a pre-shock mobility panel and weekly outcomes exist.

That portability is the larger promise. The regularity we report, behavioral spillover concentrated where routine interaction and similarity overlap, should appear wherever risk is uncertain, evolving, and socially contested, and the empirical template of a pre-shock interaction network, a lagged-peer specification, and a similarity modulator applies directly. Climate adaptation is the clearest next case: heat protection, evacuation, and air-quality avoidance plausibly diffuse along the same commuter and social networks, modulated by income and demographic similarity, and the relevant mobility data are increasingly available. Whether the mechanism reaches slower-moving domains such as household finance, where formal financial access reshapes the informal risk-sharing arrangements that link households \cite{comolaprina2023}, or vaccine uptake is an open question the design is built to test. The pandemic case makes the general point hard to avoid: communities do not meet risk in isolation, and models that assume they do, whether of disease, of climate, or of financial contagion, will misread which interventions move which populations and misallocate the attention that decides whether a society absorbs the next systemic shock or amplifies it.

\section*{Methods}

\subsection*{Data and outcome construction}

We combine three town-level data sources for 313 of the 351 Massachusetts municipalities: SafeGraph (Advan) device-flow data, weekly COVID-19 case counts from the Massachusetts Department of Public Health, and U.S. Census Bureau socioeconomic and demographic features. The 38 missing municipalities are small western and central towns absent from the SafeGraph place-level panel; their exclusion is a structural property of the device-flow source rather than an analytic choice.

The mobility data are anonymized device-level location records, originally at the census-block-group (CBG) level, aggregated to census tract and then to town using Census CBG$\to$tract and tract$\to$town crosswalks. Town is the operative unit for three reasons: COVID-19 case data are reported weekly at the town level; town aligns with public-health and municipal policy authority in Massachusetts; and place-level aggregation absorbs CBG-to-tract noise from device counts. The town panel runs weekly from early March 2020 through March 2021 (52 weeks). Boston is a single SafeGraph place; the main-text regressions treat it as one node, while the community-map figures expand it into 14 known neighborhood polygons at plot time.

Local hazard is the town-week count of weekly new cases per 10{,}000 residents, denoted $X_{i,t}$. Mortality and hospitalization counts are not used: mortality is not town-level-resolved for most of 2020, and hospitalizations conflate residence with the location of higher-capacity hospitals, biasing apparent objective risk toward urban towns regardless of behavior. Demographic features are taken at the town level: median age, household net worth, a diversity index, population density per square mile, renter share, and a seven-share race-composition vector (Hispanic, Asian, American Indian, Black, White, Mixed, Other). For the 14 Boston neighborhoods, features are population-weighted into a single Boston row to keep alignment with the SafeGraph place granularity.

The outcome variable is the within-town-demeaned median non-home dwell time. Let $D_{i,t}$ denote the median non-home dwell time of town $i$ in week $t$ (the SafeGraph published quantity, in minutes per day), and let $\bar{D}_{i}^{\mathrm{pre}}$ denote its town-specific mean over the last two pre-pandemic weeks of January 2020. The regression outcome is then
\begin{equation}
Y^{adj}_{i,t} \;=\; D_{i,t} - \bar{D}_{i}^{\mathrm{pre}},
\label{eq:outcome}
\end{equation}
in minutes per day. Demeaning by the last two pre-pandemic weeks removes time-invariant cross-town differences in baseline mobility and lets the coefficient on $X_{i,t}$ be read directly as the change in minutes per day per unit incidence change. We use minutes per day rather than log dwell time because $Y^{adj}_{i,t}$ takes both signs and zero, and the linear scale matches the unit of SafeGraph's published median dwell-time series. The panel is indexed by town $i \in \{1,\dots,313\}$ and week $t \in \{1,\dots,52\}$, giving $N_{\text{obs}} = 16{,}276$ town-week observations unless otherwise noted; missing town-weeks are dropped pairwise.

\begin{figure}[ht]
    \centering
    \includegraphics[width=\linewidth]{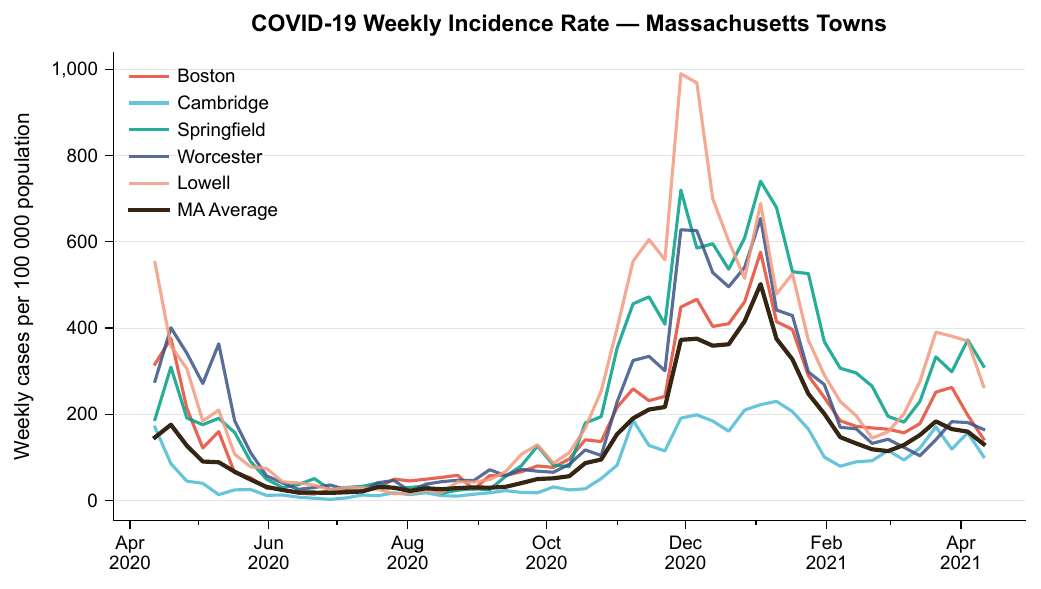}
    \caption{\textbf{Weekly COVID-19 incidence across selected Massachusetts towns and the statewide average.} Weekly new cases per $10{,}000$ residents, March 2020--April 2021. Cross-town variation in the timing and amplitude of waves provides the identifying variation in local hazard $X_{i,t}$ exploited in the regressions of §\ref{sec:methods-regression}.}
    \label{fig:weekly_incidence}
\end{figure}

\subsection*{Pre-pandemic mobility network (\texorpdfstring{$\mathrm{PreG}$}{PreG})}

The pre-pandemic mobility network $\mathrm{PreG}$ is the empirical scaffold along which the spillover regressions allow behavioral signals to propagate. Because it is constructed before any pandemic exposure, it is not a function of the pandemic-period outcome, which addresses the most direct simultaneity concern. The source is the town-to-town SafeGraph flow matrix, aggregated from device-level visits during the last two weeks of January 2020, the most recent pre-pandemic window for which clean data are available, before the first confirmed Massachusetts COVID-19 case.

From this matrix we derive the analytic default used throughout: the origin-normalized form $\mathrm{PreG}_{ij} = (\text{flow}_{i \to j}) / \sum_{k \ne i} \text{flow}_{i \to k}$, so that each row sums to one with the diagonal zeroed. For community detection the matrix is symmetrized to $\widetilde{\mathrm{PreG}} = (\mathrm{PreG} + \mathrm{PreG}^{\top}) / 2$ with the diagonal zeroed; the asymmetric origin-normalized form is retained for the regression-side spillover regressor, while the symmetrized form is used only for Louvain modularity optimization. The 313 SafeGraph-covered towns are held in a fixed canonical order that all subsequent matrices ($D$, the masks, and the regression panels) inherit, and the 38 SafeGraph-absent towns are excluded throughout. The spillover regressor is as in Eq.~\eqref{eq:s_preg}, $S^{\mathrm{PreG}}_{i,t-1} = \sum_j \mathrm{PreG}_{ij}\,Y^{adj}_{j,t-1}$, with $\mathrm{PreG}_{ii} = 0$. Five flow-matrix variants (raw, summed, origin-normalized, destination-normalized, and the four with-diagonal counterparts) are produced by the public reproduction code and described in SI~\S\ref{suppsubsec:methods-D}.

\subsection*{Community detection}\label{sec:methods-communities}

The mobility and demographic partitions used in Figs.~\ref{fig:partitions} and~\ref{fig:results_main}\textbf{c} are both obtained by Louvain modularity optimization \cite{blondel2008fast} of an appropriate town-by-town weighted graph. We chose Louvain over alternatives such as Clauset--Newman--Moore greedy agglomeration \cite{clauset2004finding}, label propagation, and spectral cuts for three reasons: it admits a tunable resolution parameter $\gamma$ that lets us report the modularity-maximizing partition rather than fix an unobservable resolution in advance; a sweep across $\gamma$ is essentially free at this graph size ($n = 313$); and it is reproducible across runs at a fixed random seed. The Clauset--Newman--Moore algorithm is also run, on the same graphs, as a baseline check (SI §\ref{suppsubsec:greedy-cd}); the two algorithms agree on the broad community structure, and Louvain attains a higher modularity at the reported resolution in both the mobility and demographic settings.

\subsubsection*{Algorithm and resolution-sweep tuning.}
Louvain is run with the \texttt{python-louvain} implementation \cite{blondel2008fast}, weight-aware, with a deterministic random seed. For each graph we sweep the resolution parameter $\gamma$ on a $20$-point grid from $0.1$ to $2.0$ in steps of $0.1$, recompute modularity $Q(\gamma)$ at every step, and adopt the partition whose modularity is maximal. The mobility sweep on $\widetilde{\mathrm{PreG}}$ has a clear global maximum at $\gamma = 0.70$ ($Q = 0.670$, $k = 7$); the demographic sweep on the RBF-kernel graph has a comparably clean maximum at $\gamma = 0.60$ ($Q = 0.859$, $k = 8$). Lower-$\gamma$ partitions collapse communities and raise modularity only mildly, while higher-$\gamma$ partitions over-fragment the network and modularity falls monotonically. We commit to these two $\gamma$ values throughout the paper; the full resolution-sweep curves (SI Fig.~\ref{fig:si_resolution_sweep}) and the corresponding greedy-modularity comparison are in SI~\S\ref{suppsubsec:greedy-cd}.

\subsubsection*{Mobility partition (main analysis).}
The mobility partition is run on $\widetilde{\mathrm{PreG}}$, the symmetrized origin-normalized pre-pandemic flow matrix with zeroed diagonal. The reported partition at $\gamma = 0.70$ has $Q = 0.670$ and $k = 7$ communities of sizes $[98, 62, 56, 40, 24, 20, 13]$. The communities are spatially coherent (Greater Boston/Northeast; Cape Cod and South Coast; Pioneer Valley; Berkshires; north-central; central; islands) but do not align with Massachusetts county boundaries, as the dashed-county overlay in Fig.~\ref{fig:partitions}\textbf{a} makes explicit.

\subsubsection*{Demographic partition.}
The demographic partition is built from a town-by-town RBF kernel on a four-feature standardized vector: median age, household net worth, diversity index, and population density. The kernel is sparsified by mutual-$k$NN with $k = 12$ to remove the dense tail of weak similarities that would otherwise dominate the modularity computation, and Louvain is then run on the resulting sparse weighted graph over the same $\gamma$-grid. The reported partition at $\gamma = 0.60$ has $Q = 0.859$ and $k = 8$. We use this four-feature categorical partition, rather than the richer continuous demographic-similarity matrix $D$ of SI~\S\ref{suppsubsec:methods-D}, specifically for the four-way decomposition (Fig.~\ref{fig:results_main}\textbf{c}), because the four-way crossing requires a partition on each axis. The matrix $D$ enters the regressions as a continuous modulator (SI~\S\ref{suppsec:demographic-robustness}), and the two are complementary tests of the same structural claim. On the 313-town intersection of the two partitions, $\mathrm{ARI} = 0.082$ and $\mathrm{NMI} = 0.144$: the partitions index nearly independent dimensions of inter-town structure (Fig.~\ref{fig:partitions}\textbf{f}), which is the formal precondition for the four-way decomposition to carry non-redundant information. Implementation details and the corresponding greedy-modularity baseline are in SI~\S\ref{suppsubsec:greedy-cd}.

\subsection*{Regression specification and identification}\label{sec:methods-regression}

All regressions are two-way fixed-effects panel models on $N = 16{,}276$ town-weeks (313 towns $\times$ 52 weeks). The pooled headline specification (Table~\ref{tab:results_main}, columns 1--2) is
\begin{equation}
Y^{adj}_{i,t} \;=\; \mu_{i} + \lambda_{t} + \alpha\,X_{i,t} + \beta\,S^{\mathrm{PreG}}_{i,t-1} + \gamma\,Y^{adj}_{i,t-1} + \varepsilon_{i,t},
\label{eq:headline}
\end{equation}
where $\mu_{i}$ are town fixed effects, $\lambda_{t}$ are week fixed effects, and $\varepsilon_{i,t}$ is the disturbance. The three structural coefficients $(\alpha, \beta, \gamma)$ correspond, respectively, to the response to local objective risk, the inter-town behavioral spillover along $\mathrm{PreG}$, and within-town behavioral inertia; this $(\alpha, \beta, \gamma)$ labelling is used consistently throughout the Results and figures. Because the unit of analysis is the town, $\gamma$ does not isolate a single individual-level mechanism: at this aggregation, the lagged-outcome channel from $Y^{adj}_{i,t-1}$ to $Y^{adj}_{i,t}$ combines residents' own-behavior persistence with within-town peer reinforcement, both of which propagate intra-town from week to week. Inter-town peer effects, by contrast, enter only through $\beta$ and its mask-decomposed variants. Standard errors are clustered at the town level to allow within-town serial dependence and arbitrary intra-town heteroskedasticity.

The specification has several variants. The mobility-decomposed columns (Table~\ref{tab:results_main}, cols.~3--4) replace $\beta\,S^{\mathrm{PreG}}_{i,t-1}$ with $\beta^{w}\,S^{\mathrm{PreG},w}_{i,t-1} + \beta^{b}\,S^{\mathrm{PreG},b}_{i,t-1}$, where the within/across split is the Hadamard mask of $\mathrm{PreG}$ by the mobility-Louvain partition. The demographic-decomposed columns (cols.~5--6) and the joint four-mask and continuous-modulator robustness specifications are reported in detail in SI §\ref{suppsec:demographic-robustness}.

Identification rests on breaking Manski's reflection problem \cite{manski1993identification}, which arises when an own outcome and the contemporaneous peer mean enter the same regression and are mutually determined: $Y_{i,t} = \beta\,\mathbb{E}[Y_{j,t}\mid j \sim i] + \dots$ has no separately identified $\beta$ because the conditional mean is itself a function of $Y_{i,t}$. Two design choices break this here. First, the peer signal $S^{\mathrm{PreG}}_{i,t-1}$ is the \emph{lagged} network-weighted outcome rather than the contemporaneous one, so the simultaneity that defeats $t = t$ specifications is replaced by a one-step prediction in which week-$t$ behavior of $i$ cannot mechanically drive week-$(t{-}1)$ behavior of $j$. Second, the network $\mathrm{PreG}$ is fixed in January 2020, before any case in Massachusetts, so the regressor matrix is pre-determined with respect to the pandemic-period outcome process; in the language of Bramoull\'e, Djebbari and Fortin \cite{bramoulle2009identification}, $\mathrm{PreG}$ is a transitive but pre-pandemic instrument-like structure on which the lagged-peer interaction is constructed. The design is therefore identified up to the standard residual concern of correlated unobserved shocks, addressed next.

Town fixed effects $\mu_{i}$ absorb all time-invariant municipal characteristics (geography, demographic baseline, institutional capacity), and week fixed effects $\lambda_{t}$ absorb common shocks (state-level policy stringency, national news, school-calendar effects). The residual concern is spatially patterned, time-varying unobservables that travel along the same edges as $\mathrm{PreG}$ \cite{bramoulle2009identification}, for example a regional outbreak that simultaneously shocks behavior in connected towns through information rather than through peer adjustment. We address this in two ways. First, a network-exposure-to-cases falsification replaces $S^{\mathrm{PreG}}_{i,t-1}$ with $S^{\mathrm{PreG}\cdot X}_{i,t-1} = \sum_j \mathrm{PreG}_{ij}\,X_{j,t-1}$ and yields a coefficient statistically indistinguishable from zero; when both terms enter side by side, the network-exposure-to-cases coefficient remains insignificant while $S^{\mathrm{PreG}}_{i,t-1}$ retains its sign and magnitude (SI Table~\ref{tab:models2_with_excluded_towns}). A pure information-shock alternative should load on $X$-exposure rather than $Y$-exposure, and it does not. Second, a geographic-proximity placebo network replaces $\mathrm{PreG}$ with an inverse-distance or shared-county adjacency matrix; because this structure encodes spatial correlation without routine interaction, a channel specific to the mobility network predicts an attenuated placebo coefficient, isolating the mobility-specific content of the spillover from generic spatial autocorrelation.

Endogenous group formation is not a threat in this design. Because $\mathrm{PreG}$ is constructed before the pandemic, peer-group composition is exogenous to the outcome shock, so the standard concern that agents select into groups on the same unobservables that drive outcomes \cite{jackson2008social} does not apply: towns that share heavy pre-pandemic flow did so for reasons set years before COVID-19, such as commuting, retail catchment, and social ties.

The within-town channel $\gamma$ and the across-town channel $\beta$ are separately identified because they load on different lagged objects: $\gamma$ on $Y^{adj}_{i,t-1}$, the focal town's own lagged outcome, and $\beta$ on $\sum_j \mathrm{PreG}_{ij}\,Y^{adj}_{j,t-1}$ with $\mathrm{PreG}_{ii} = 0$, a network-weighted sum over \emph{other} towns. The two regressors are correlated at the town-week level, since towns with high lagged own outcomes tend to sit in communities whose peers also had high lagged outcomes, but they are not collinear, and the FE+AR(1) specifications in Table~\ref{tab:results_main} identify both with non-trivial standard errors. Because the two are positively correlated and $\gamma > 0$, omitting the own lag (column~1) biases the peer coefficient upward by $\gamma\,\delta$, where $\delta$ is the within-fixed-effects regression slope of $Y^{adj}_{i,t-1}$ on $S^{\mathrm{PreG}}_{i,t-1}$; this is why $\hat\beta$ falls from $0.51$ to $0.23$ once the lag enters. The FE+AR(1) estimate is therefore the partialed across-town response and the value that serves as the lower bound on social influence, while the larger static coefficient is an interior point of the $[\hat\beta,\, \hat\beta + \hat\gamma]$ interval rather than a bound on it. The substantive interpretation respects this split: $\gamma$ captures the joint persistence-plus-within-town-reinforcement that operates intra-town from week to week, and $\beta$ captures the across-town peer effect that the network can carry but the lagged-own term cannot.

Regressions are estimated with \texttt{statsmodels} (PanelOLS with two-way fixed effects and clustered standard errors). Community detection uses \texttt{python-louvain}, with \texttt{networkx} for graph manipulation; robust distances use the \texttt{MinCovDet} estimator in \texttt{scikit-learn}; and spatial layers use \texttt{geopandas} and \texttt{pyproj} (EPSG:26986). The full processing pipeline and the corresponding scripts are available in the public reproduction repository.

\subsection*{Hadamard masks and decompositions}\label{sec:methods-masks}

The within/across mask used in the main-text within-mobility-community decomposition and the four-mask cross-decomposition used in the SI demographic robustness are defined as element-wise products of $\mathrm{PreG}$ with $\{0,1\}$ indicator matrices.

For a partition $\Pi = \{C_1, \dots, C_K\}$ of the 313 towns, the within-mask and across-mask are $M^{w}_{ij}(\Pi) = \mathbf{1}[\Pi(i) = \Pi(j),\, i \neq j]$ and $M^{b}_{ij}(\Pi) = \mathbf{1}[\Pi(i) \neq \Pi(j)]$, so that $\mathrm{PreG}^{w}(\Pi) = \mathrm{PreG} \odot M^{w}(\Pi)$ and $\mathrm{PreG}^{b}(\Pi) = \mathrm{PreG} \odot M^{b}(\Pi)$, where $\odot$ is the elementwise (Hadamard) product. These are not row-renormalized after masking, so the masked-spillover coefficient is on the same scale as the pooled $S^{\mathrm{PreG}}$.

Crossing the mobility partition $\Pi^{M}$ with the demographic partition $\Pi^{D}$ yields four masks,
\begin{align*}
\mathrm{PreG}^{mm} &= \mathrm{PreG} \odot M^{w}(\Pi^{M}) \odot M^{w}(\Pi^{D}), &
\mathrm{PreG}^{ma} &= \mathrm{PreG} \odot M^{w}(\Pi^{M}) \odot M^{b}(\Pi^{D}), \\
\mathrm{PreG}^{am} &= \mathrm{PreG} \odot M^{b}(\Pi^{M}) \odot M^{w}(\Pi^{D}), &
\mathrm{PreG}^{aa} &= \mathrm{PreG} \odot M^{b}(\Pi^{M}) \odot M^{b}(\Pi^{D}),
\end{align*}
which sum to $\mathrm{PreG}$ off-diagonal. Each spillover regressor is then $S^{kl}_{i,t-1} = \sum_j \mathrm{PreG}^{kl}_{ij}\,Y^{adj}_{j,t-1}$ for $kl \in \{mm, ma, am, aa\}$. Finally, the matrix $D$ supports an alternative continuous-modulator spillover $S^{\mathrm{PreG}\odot D}_{i,t-1}$ used in the SI demographic robustness; its formal definition is given in SI §\ref{suppsubsec:methods-D}.

\bmhead{Acknowledgements}
This material is based upon work supported by the National Science Foundation under Grant No.~DMS-2421289 (IHBEM: \emph{No One Lives in a Bubble: Incorporating Group Dynamics into Epidemic Models}), awarded to Northeastern University. Any opinions, findings, and conclusions or recommendations expressed in this material are those of the authors and do not necessarily reflect the views of the National Science Foundation.

\bibliography{bibliography}

\clearpage
\setcounter{section}{0}
\setcounter{subsection}{0}
\setcounter{subsubsection}{0}
\setcounter{table}{0}
\setcounter{figure}{0}
\setcounter{equation}{0}
\renewcommand{\thesection}{S\arabic{section}}
\renewcommand{\thetable}{S\arabic{table}}
\renewcommand{\thefigure}{S\arabic{figure}}
\renewcommand{\theequation}{S\arabic{equation}}

\section*{Supplementary Information}

\section*{Supplementary Methods}

This section collects the construction details deferred from the main-text Methods: the continuous demographic-similarity matrix $D$, the community-detection resolution sweep, and the greedy-modularity baseline. As context for the outcome variable, Fig.~\ref{fig:non_home_dwell_time_ma_vs_boston} contrasts the statewide non-home dwell time with that of Boston over the first pandemic year, which is why Boston, a single SafeGraph place, enters the regressions as one node and is expanded into its constituent neighborhoods only for the community maps.

\IfFileExists{figs/SI/non_home_dwell_time_ma_vs_boston.pdf}{%
\begin{figure}[htbp]
    \centering
    \includegraphics[width=\linewidth]{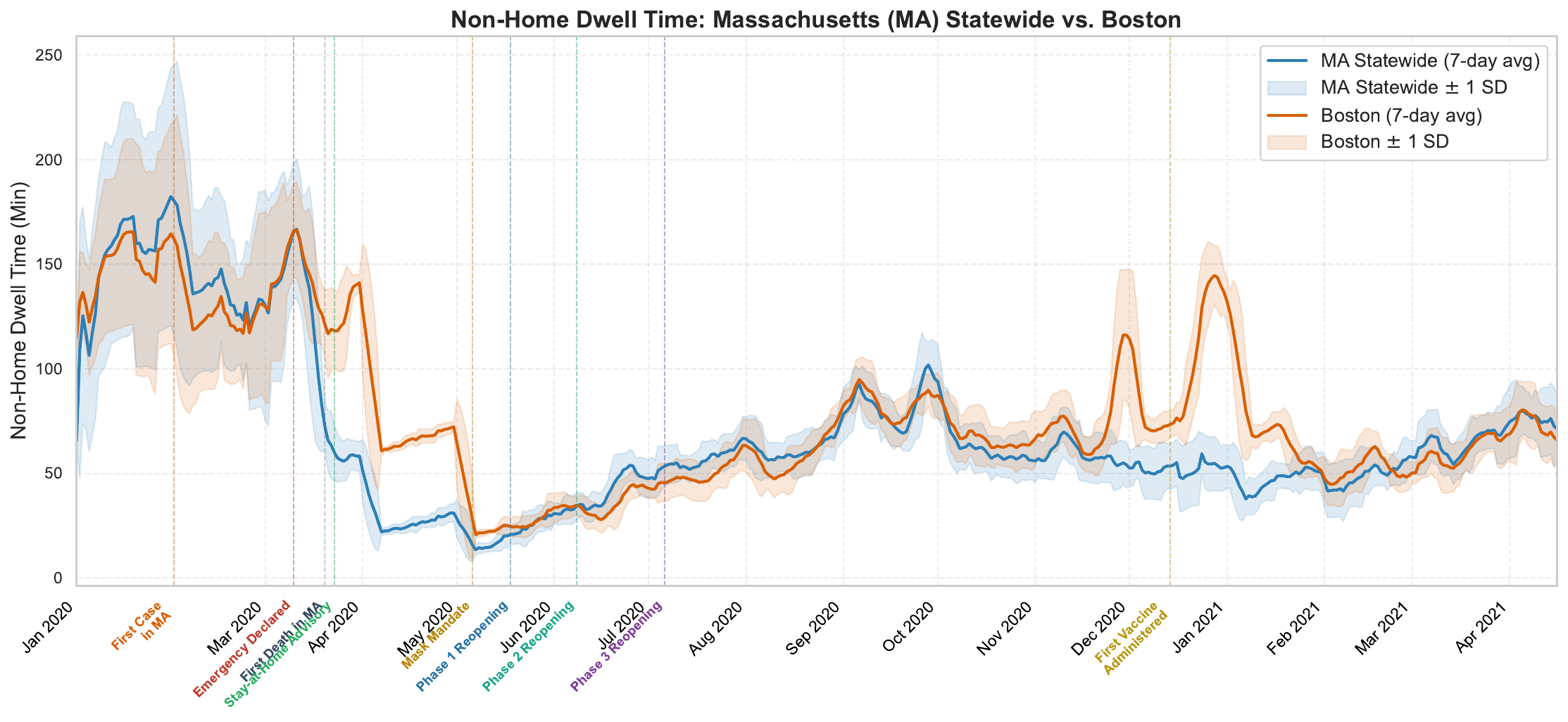}
    \caption{\textbf{Non-home dwell time, Massachusetts statewide average versus Boston.} Weekly median non-home dwell time over the first pandemic year for the statewide average and for Boston, shown together for context. Boston is a single SafeGraph place and enters the regressions as one node.}
    \label{fig:non_home_dwell_time_ma_vs_boston}
\end{figure}%
}{}

\subsection*{Demographic similarity matrix \texorpdfstring{$D$}{D}}\label{suppsubsec:methods-D}

The continuous demographic-similarity matrix $D \in \mathbb{R}^{313 \times 313}$ underpins the robustness analyses for the demographic decomposition in §\ref{suppsec:demographic-robustness}. It is built as a principled distance for a heterogeneous feature set that has both a continuous block and a compositional block, which are combined and then mapped to a similarity.

The continuous block uses the robust Mahalanobis distance on median age, household net worth, diversity index, population density, and renter share, with the precision matrix taken from a Minimum Covariance Determinant (MCD) fit \cite{rousseeuw1999fast}. Because the MCD estimator is affine-equivariant, the median-imputed features are passed in without prior $z$-scoring; a non-robust standardization would itself be distorted by the outliers the MCD is meant to discount. The compositional block handles the seven-share race vector through the centered log-ratio transform $\mathrm{clr}(\mathbf{p}) = \log\mathbf{p} - \overline{\log\mathbf{p}}$, applied after a small-value replacement for zero shares; pairwise Euclidean distance in clr space is the Aitchison distance \cite{aitchison1986statistical}.

Each block is scaled by its lower-triangle median, the two are weighted equally and summed, and the combined distance is mapped through an RBF kernel, $D_{ij} = \exp\!\big(- d^{2,\text{combined}}_{ij} / (2\sigma^2)\big)$, with $\sigma = \sqrt{\mathrm{median}(d^{2,\text{combined}}) / 2}$ set by the median heuristic \cite{garreau2017median}. The diagonal is zeroed, the matrix is symmetric by construction, and $D_{ij} \in (0, 1]$, with a mean off-diagonal value near $0.39$. The 313 towns follow the same canonical order as $\mathrm{PreG}$.

The matrix supports a continuous-modulator alternative to the binary demographic-Louvain partition,
\begin{equation}
S^{\mathrm{PreG}\odot D}_{i,t-1} \;=\; \sum_{j} \mathrm{PreG}_{ij}\, D_{ij}\, Y^{adj}_{j,t-1},
\label{eq:s_pregxd_SI}
\end{equation}
with coefficient $\beta^{\odot}$. Because $\mathrm{PreG}\odot D$ is not row-renormalized after modulation, $\beta^{\odot}$ sits on a different scale from $\beta$ and is read through its predicted impact, $\hat\beta^{\odot}\times\mathrm{sd}(S^{\mathrm{PreG}\odot D})$, rather than as a magnitude directly comparable to $\hat\beta$. The four-feature variant used for the demographic-Louvain partition differs from this five-feature block only in the renter-share dimension.

\subsection*{Community detection: resolution-sweep tuning and greedy-modularity baseline}\label{suppsubsec:greedy-cd}

The mobility and demographic Louvain partitions reported in the main text are the modularity-maximizing partitions of a resolution sweep over $\gamma \in [0.1, 2.0]$. The sweep curves appear in Fig.~\ref{fig:si_resolution_sweep}. The mobility curve, run on the symmetrized origin-normalized $\widetilde{\mathrm{PreG}}$, reaches its global maximum at $\gamma = 0.70$ ($Q = 0.670$, $k = 7$); the demographic curve, run on the RBF-kernel graph of the four-feature standardized vector sparsified by mutual-$k$NN at $k = 12$, reaches its maximum at $\gamma = 0.60$ ($Q = 0.859$, $k = 8$). Both curves fall away monotonically at high $\gamma$ as the algorithm over-fragments the network.

\IfFileExists{figs/SI/ma_communities_louvain_resolution_sweep.png}{%
\begin{figure}[!htbp]
    \centering
    \begin{minipage}{0.49\linewidth}\centering
        \includegraphics[width=\linewidth]{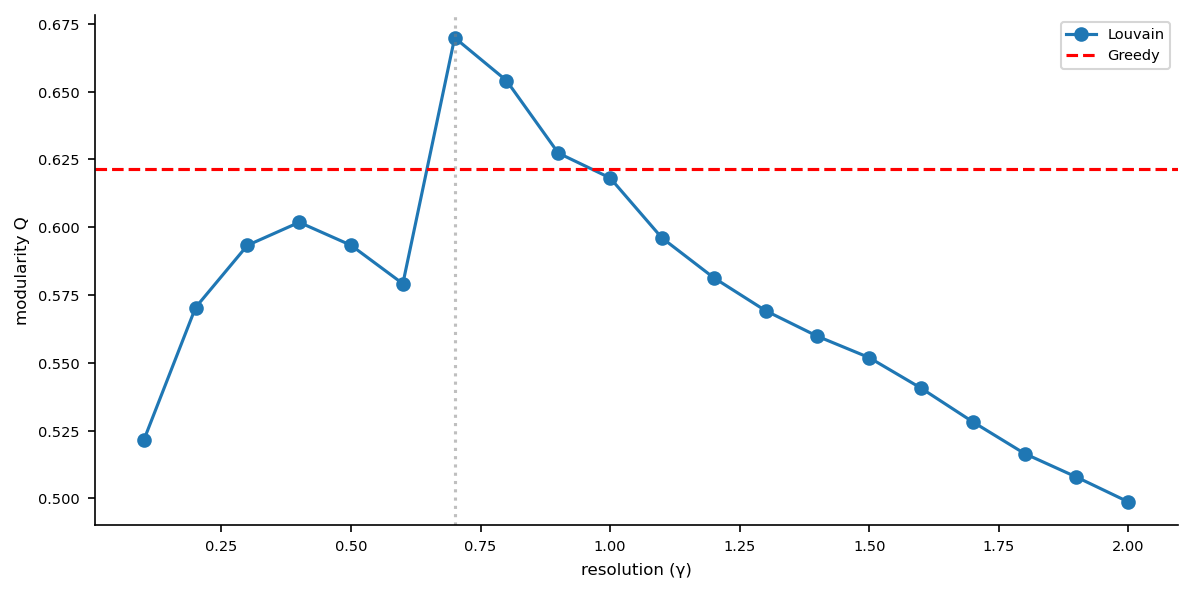}
    \end{minipage}\hfill
    \begin{minipage}{0.49\linewidth}\centering
        \includegraphics[width=\linewidth]{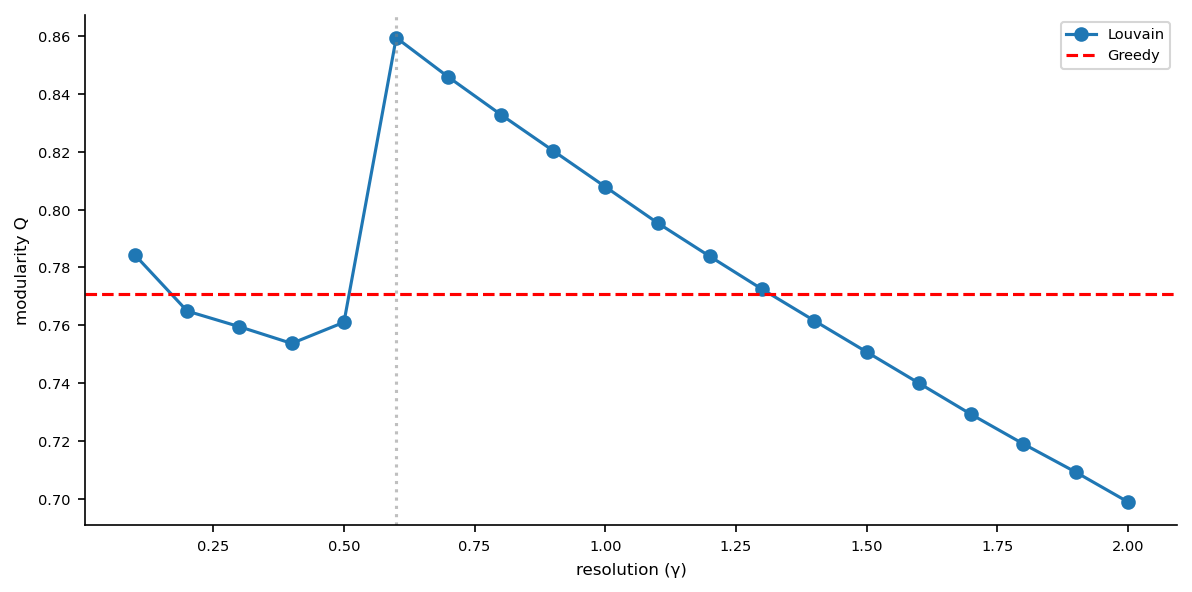}
    \end{minipage}
    \caption{\textbf{Louvain resolution-sweep tuning for the mobility (left) and demographic (right) partitions.} Modularity $Q$ as a function of the Louvain resolution parameter $\gamma$ on the $20$-point grid $\gamma \in \{0.1, 0.2, \ldots, 2.0\}$ (random seed $42$). Dotted vertical lines mark the reported $\gamma$ at the modularity-maximizing partition: $\gamma^{\star}_{\mathrm{mob}} = 0.70$ ($Q = 0.670$, $k = 7$) and $\gamma^{\star}_{\mathrm{demo}} = 0.60$ ($Q = 0.859$, $k = 8$). The horizontal dashed reference line on each panel reports the Clauset--Newman--Moore greedy-modularity benchmark on the same graph.}
    \label{fig:si_resolution_sweep}
\end{figure}%
}{%
}

\subsubsection*{Greedy-modularity baseline.}
We re-run community detection on the same two graphs with the Clauset--Newman--Moore greedy-agglomeration algorithm \cite{clauset2004finding}, the most common modularity-optimizing alternative to Louvain at this graph size and one with no resolution parameter. On $\widetilde{\mathrm{PreG}}$ greedy modularity recovers $k = 7$ mobility communities of sizes $[71, 67, 62, 41, 29, 24, 19]$ at $Q = 0.621$, and on the RBF demographic graph it recovers $k = 6$ demographic communities at $Q = 0.771$. Louvain attains a strictly higher modularity in both cases ($+0.05$ on mobility, $+0.09$ on demographic), and the partition contours largely agree: the seven greedy mobility communities are geographically the same Greater Boston, Cape Cod, Pioneer Valley, Berkshires, central, north-central, and islands clusters that Louvain recovers. The greedy maps are shown in Fig.~\ref{fig:si_greedy_maps}.

\IfFileExists{figs/SI/ma_communities_greedy.pdf}{%
\begin{figure}[!htbp]
    \centering
    \begin{minipage}{0.49\linewidth}\centering
        \includegraphics[width=\linewidth]{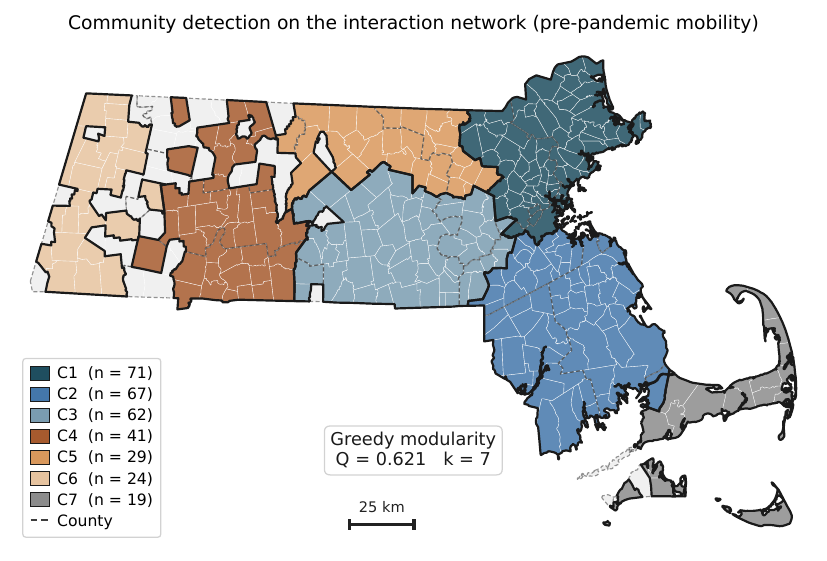}
    \end{minipage}\hfill
    \begin{minipage}{0.49\linewidth}\centering
        \includegraphics[width=\linewidth]{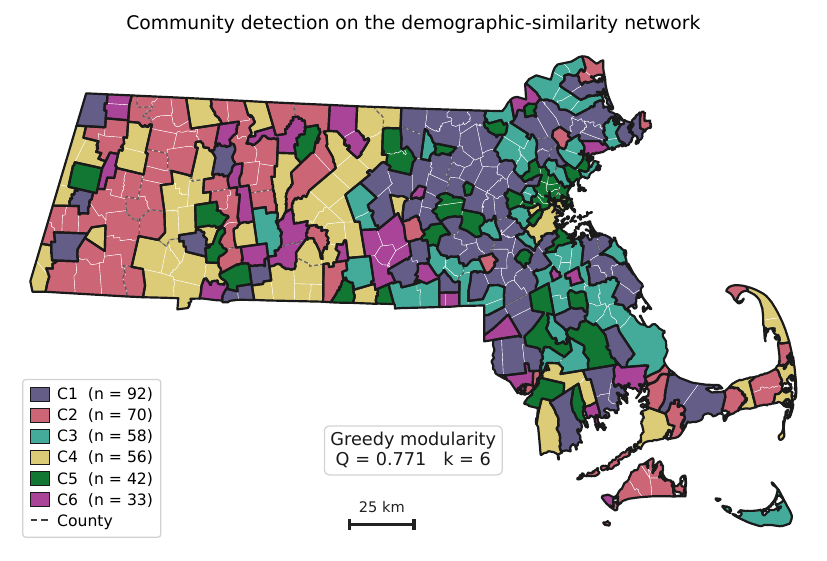}
    \end{minipage}
    \caption{\textbf{Clauset--Newman--Moore greedy-modularity partitions on the mobility (left) and demographic (right) graphs.} Town polygons are colored by greedy community; thick outlines mark community boundaries; dashed grey lines on the mobility panel mark Massachusetts county boundaries. Greedy recovers $k = 7$ mobility communities ($Q = 0.621$) and $k = 6$ demographic communities ($Q = 0.771$), both at lower modularity than the Louvain partitions reported in main-text Fig.~\ref{fig:partitions}\textbf{a},\textbf{c}.}
    \label{fig:si_greedy_maps}
\end{figure}%
}{%
}

\subsubsection*{Why Louvain.}
We adopt Louvain rather than the greedy baseline for three reasons. First, the modularity gap is small but not negligible: $0.05$ on the mobility graph and $0.09$ on the demographic graph correspond to a noticeably tighter community structure. Second, the resolution parameter $\gamma$ gives a principled handle for reporting the best partition rather than committing to an unobservable resolution before seeing the data, and the modularity-maximizing $\gamma$ is reproducible and survives small perturbations of the graph. Third, the close agreement between the two partitions documented above gives no reason to expect the substantive conclusions to depend on the choice between Louvain and greedy agglomeration.

\section*{Supplementary Results}
% Include additional results, figures, tables to support main text

\subsection*{Detailed regression specifications underlying the main spillover table}\label{suppsubsec:detailed_regressions}

This subsection reports the full set of regression specifications that underlie main-text Table~\ref{tab:results_main}. The merged main-text table extracts, from each of three Hadamard-masking schemes (pooled, mobility within / across, demographic within / across), only the two columns most relevant to the headline narrative, one without and one with the lagged dependent variable. Tables~\ref{tab:si_main_six_models}--\ref{tab:si_demographic_partition} report the same three specifications in full, including auxiliary columns that re-introduce the unsplit network spillover $S^{\mathrm{PreG}}_{i,t-1}$ alongside the masked components. These columns show which masked component absorbs the network channel and how the unsplit term behaves once a partition is included, contrasts that motivate the main-text prose but are not visible in the merged display.

\begin{table}[!htbp]
\centering
\caption{\textbf{Pooled spillover and mobility-community decomposition (full specification).} This table is the unabbreviated version of the pooled and mobility-decomposed blocks of main-text Table~\ref{tab:results_main}. Outcome: $Y^{adj}_{i,t}$, weekly median non-home dwell time minus the town's pre-pandemic baseline (minutes per day). $X_{i,t}$ is weekly cases per $10{,}000$. $S^{\mathrm{PreG}}_{i,t-1}$ is the network-weighted lagged behavior of pre-pandemic mobility peers; $S^{\mathrm{PreG},w}$ and $S^{\mathrm{PreG},b}$ Hadamard-mask $\mathrm{PreG}$ to within- and across-community pairs of the mobility-Louvain partition. All columns include town and week fixed effects; standard errors clustered at the town level in parentheses. $^{*}p<0.10$, $^{**}p<0.05$, $^{***}p<0.01$.}
\label{tab:si_main_six_models}
\footnotesize
\setlength{\tabcolsep}{0pt}
\begin{tabular*}{\linewidth}{@{\extracolsep{\fill}} l ccc}
\toprule
 & \multicolumn{3}{c}{\textit{Dep.\ var.: $Y^{adj}_{i,t}$ (min/day)}} \\
\cmidrule(lr){2-4}
 & (1) & (2) & (3) \\
 & FE, pooled & FE + AR(1), pooled & FE, mobility within / across \\
\midrule
$Y^{adj}_{i,t-1}$ & & $0.457^{***}$ & \\
 & & $(0.046)$ & \\[3pt]

$X_{i,t}$ (per 10k) & $-0.110^{***}$ & $-0.069^{***}$ & $-0.109^{***}$ \\
 & $(0.042)$ & $(0.025)$ & $(0.042)$ \\[3pt]

$S^{\mathrm{PreG}}_{i,t-1}$ & $0.512^{***}$ & $0.227^{***}$ & \\
 & $(0.191)$ & $(0.088)$ & \\[3pt]

$S^{\mathrm{PreG},w}_{i,t-1}$ & & & $0.528^{***}$ \\
 & & & $(0.200)$ \\[3pt]

$S^{\mathrm{PreG},b}_{i,t-1}$ & & & $-0.083$ \\
 & & & $(0.294)$ \\
\midrule
$R^{2}$            & 0.772 & 0.822 & 0.772 \\
$N$ (town-weeks)   & 16{,}276 & 16{,}276 & 16{,}276 \\
Town FE            & Yes & Yes & Yes \\
Week FE            & Yes & Yes & Yes \\
\bottomrule
\end{tabular*}
\vspace{4pt}
{\scriptsize \textit{Notes.} Column 1 is the two-way fixed-effects model with the unsplit network-mediated spillover and local hazard. Column 2 adds the lagged dependent variable. Column 3 splits the spillover into within- and across-mobility-community components using the Louvain partition of $\mathrm{PreG}$ ($k = 7$ communities, $Q = 0.67$).}
\end{table}

\begin{table}[!htbp]
\centering
\caption{\textbf{Within / across decomposition of the network spillover by the mobility-Louvain partition (auxiliary specifications).} Outcome and notation as in Table~\ref{tab:si_main_six_models}. Columns~1 and~3 reproduce the mobility-decomposed columns of main-text Table~\ref{tab:results_main} (columns~3 and~4). Columns~2 and~4 add the unsplit network spillover $S^{\mathrm{PreG}}_{i,t-1}$ alongside the within and across components. The within-community term remains significant in every specification while the across-community term is statistically indistinguishable from zero throughout. The unsplit term loses significance once the partition is included: the within-community pairs absorb the network channel. Mobility-Louvain partition: Fig.~\ref{fig:partitions}a, $k = 7$, $Q = 0.67$. $^{*}p<0.10$, $^{**}p<0.05$, $^{***}p<0.01$.}
\label{tab:si_mobility_partition}
\footnotesize
\setlength{\tabcolsep}{0pt}
\begin{tabular*}{\linewidth}{@{\extracolsep{\fill}} l cccc}
\toprule
 & \multicolumn{4}{c}{\textit{Dep.\ var.: $Y^{adj}_{i,t}$ (min/day)}} \\
\cmidrule(lr){2-5}
 & (1) & (2) & (3) & (4) \\
 & FE, w/b & FE, w/b + $S^{\mathrm{PreG}}$ & FE+AR, w/b & FE+AR, w/b + $S^{\mathrm{PreG}}$ \\
\midrule
$Y^{adj}_{i,t-1}$ & & & $0.456^{***}$ & $0.456^{***}$ \\
 & & & $(0.046)$ & $(0.046)$ \\[3pt]

$X_{i,t}$ (per 10k) & $-0.109^{***}$ & $-0.109^{***}$ & $-0.068^{***}$ & $-0.068^{***}$ \\
 & $(0.042)$ & $(0.042)$ & $(0.025)$ & $(0.025)$ \\[3pt]

$S^{\mathrm{PreG}}_{i,t-1}$ & & $0.148$ & & $0.054$ \\
 & & $(0.128)$ & & $(0.065)$ \\[3pt]

$S^{\mathrm{PreG},w}_{i,t-1}$ & $0.528^{***}$ & $0.379^{**}$ & $0.235^{**}$ & $0.182^{**}$ \\
 & $(0.200)$ & $(0.152)$ & $(0.092)$ & $(0.073)$ \\[3pt]

$S^{\mathrm{PreG},b}_{i,t-1}$ & $-0.083$ & $-0.231$ & $-0.074$ & $-0.128$ \\
 & $(0.294)$ & $(0.196)$ & $(0.157)$ & $(0.104)$ \\
\midrule
$R^{2}$            & 0.772 & 0.772 & 0.822 & 0.822 \\
$N$ (town-weeks)   & 16{,}276 & 16{,}276 & 16{,}276 & 16{,}276 \\
Town FE            & Yes & Yes & Yes & Yes \\
Week FE            & Yes & Yes & Yes & Yes \\
\bottomrule
\end{tabular*}
\vspace{4pt}
{\scriptsize \textit{Notes.} Columns 2 and 4 add the unsplit network spillover $S^{\mathrm{PreG}}_{i,t-1}$ alongside the within and across components. The within-community term remains significant in every specification while the across-community term is statistically indistinguishable from zero throughout. The unsplit term loses significance once the partition is included: the within-community pairs absorb the channel.}
\end{table}

\begin{table}[!htbp]
\centering
\caption{\textbf{Within / across decomposition of the network spillover by the demographic-Louvain partition (auxiliary specifications).} Same specifications and conventions as Table~\ref{tab:si_mobility_partition}, but with the partition derived from a demographic-similarity graph (Fig.~\ref{fig:partitions}c; $k = 8$, $Q = 0.86$). Columns~1 and~3 reproduce the demographic-decomposed columns of main-text Table~\ref{tab:results_main} (columns~5 and~6). The contrast with Table~\ref{tab:si_mobility_partition} is sharp: under the demographic partition, the unsplit $S^{\mathrm{PreG}}_{i,t-1}$ remains positive and significant when added alongside the within and across components, while the across-demographic term collapses, so demographic similarity tracks behavior partially but does not, on its own, carve up the network channel. $^{*}p<0.10$, $^{**}p<0.05$, $^{***}p<0.01$.}
\label{tab:si_demographic_partition}
\footnotesize
\setlength{\tabcolsep}{0pt}
\begin{tabular*}{\linewidth}{@{\extracolsep{\fill}} l cccc}
\toprule
 & \multicolumn{4}{c}{\textit{Dep.\ var.: $Y^{adj}_{i,t}$ (min/day)}} \\
\cmidrule(lr){2-5}
 & (1) & (2) & (3) & (4) \\
 & FE, w/b & FE, w/b + $S^{\mathrm{PreG}}$ & FE+AR, w/b & FE+AR, w/b + $S^{\mathrm{PreG}}$ \\
\midrule
$Y^{adj}_{i,t-1}$ & & & $0.455^{***}$ & $0.455^{***}$ \\
 & & & $(0.047)$ & $(0.047)$ \\[3pt]

$X_{i,t}$ (per 10k) & $-0.112^{***}$ & $-0.112^{***}$ & $-0.069^{***}$ & $-0.069^{***}$ \\
 & $(0.042)$ & $(0.042)$ & $(0.024)$ & $(0.024)$ \\[3pt]

$S^{\mathrm{PreG}}_{i,t-1}$ & & $0.617^{***}$ & & $0.259^{***}$ \\
 & & $(0.161)$ & & $(0.080)$ \\[3pt]

$S^{\mathrm{Demo},w}_{i,t-1}$ & $1.446^{***}$ & $0.829^{***}$ & $0.591^{***}$ & $0.332^{***}$ \\
 & $(0.416)$ & $(0.269)$ & $(0.203)$ & $(0.128)$ \\[3pt]

$S^{\mathrm{Demo},b}_{i,t-1}$ & $0.404^{**}$ & $-0.212$ & $0.187^{**}$ & $-0.073$ \\
 & $(0.158)$ & $(0.151)$ & $(0.075)$ & $(0.068)$ \\
\midrule
$R^{2}$            & 0.773 & 0.773 & 0.822 & 0.822 \\
$N$ (town-weeks)   & 16{,}276 & 16{,}276 & 16{,}276 & 16{,}276 \\
Town FE            & Yes & Yes & Yes & Yes \\
Week FE            & Yes & Yes & Yes & Yes \\
\bottomrule
\end{tabular*}
\vspace{4pt}
{\scriptsize \textit{Notes.} The demographic-Louvain partition does not separate the spillover the way the mobility partition does: in column~1 both within- and across-community demographic spillovers are positive and significant. Once the unsplit network spillover $S^{\mathrm{PreG}}_{i,t-1}$ is included (columns~2,\,4), the across-demographic term collapses to insignificance while the unsplit $S^{\mathrm{PreG}}$ stays positive and significant, a sharp contrast with Table~\ref{tab:si_mobility_partition}, where it does not. Demographic similarity therefore tracks behavior partially, but does not, on its own, carve up the network channel.}
\end{table}

\subsection*{Behavioral co-movement under the demographic-community ordering}\label{suppsubsec:behavioral_similarity_demographic}

The block-diagonal heatmap in main-text Fig.~\ref{fig:partitions}\textbf{e}, with rows ordered by the mobility-Louvain partition, is reproduced here under an alternative ordering by the demographic-Louvain partition (Fig.~\ref{fig:si_behavioral_similarity_demographic}). The construction is identical, namely pairwise Pearson correlation of week-detrended $Y^{\mathrm{adj}}$ trajectories across the 313 towns with within-block hierarchical-cluster reordering for visual coherence; only the row and column ordering differs. The block structure that emerges so cleanly under the mobility ordering is markedly weaker under the demographic ordering: within-demographic-community correlations are only modestly elevated above across-community correlations, and the visual signature that distinguishes ``inside'' from ``outside'' a community largely vanishes. This is consistent with the regression evidence in main-text Table~\ref{tab:results_main} (columns~5--6) and Table~\ref{tab:si_demographic_partition}: demographic similarity tracks behavioral co-movement only partially and does not, on its own, carve up the network channel.

\IfFileExists{figs/SI/behavioral_similarity_demographic.pdf}{%
\begin{figure}[!htbp]
    \centering
    \includegraphics[width=0.85\linewidth]{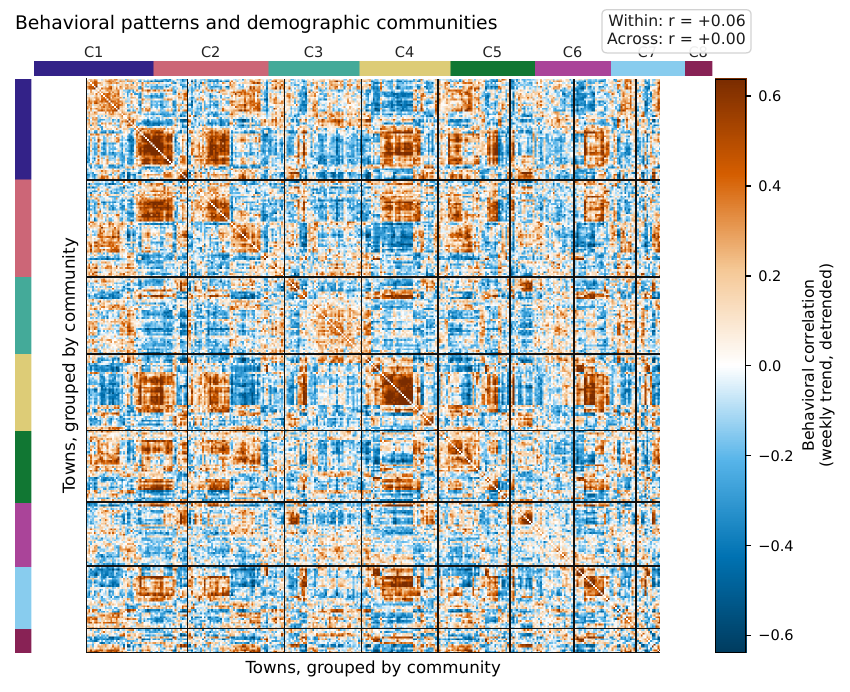}
    \caption{\textbf{Behavioral co-movement, ordered by demographic communities.} Same construction as main-text Fig.~\ref{fig:partitions}\textbf{e} but with rows and columns ordered by the demographic-Louvain partition rather than the mobility-Louvain partition. The block-diagonal structure is markedly weaker, consistent with demographic similarity acting as a modulator of the mobility-network channel rather than an independent channel of behavioral co-movement.}
    \label{fig:si_behavioral_similarity_demographic}
\end{figure}%
}{%
}

\section*{Robustness tests}\label{suppsec:robustness}
\subsection*{Excluding high-population municipalities}\label{suppsubsec:robustness-exclude-towns}

\begin{figure}[htbp]
\centering
\includegraphics[width=\textwidth]{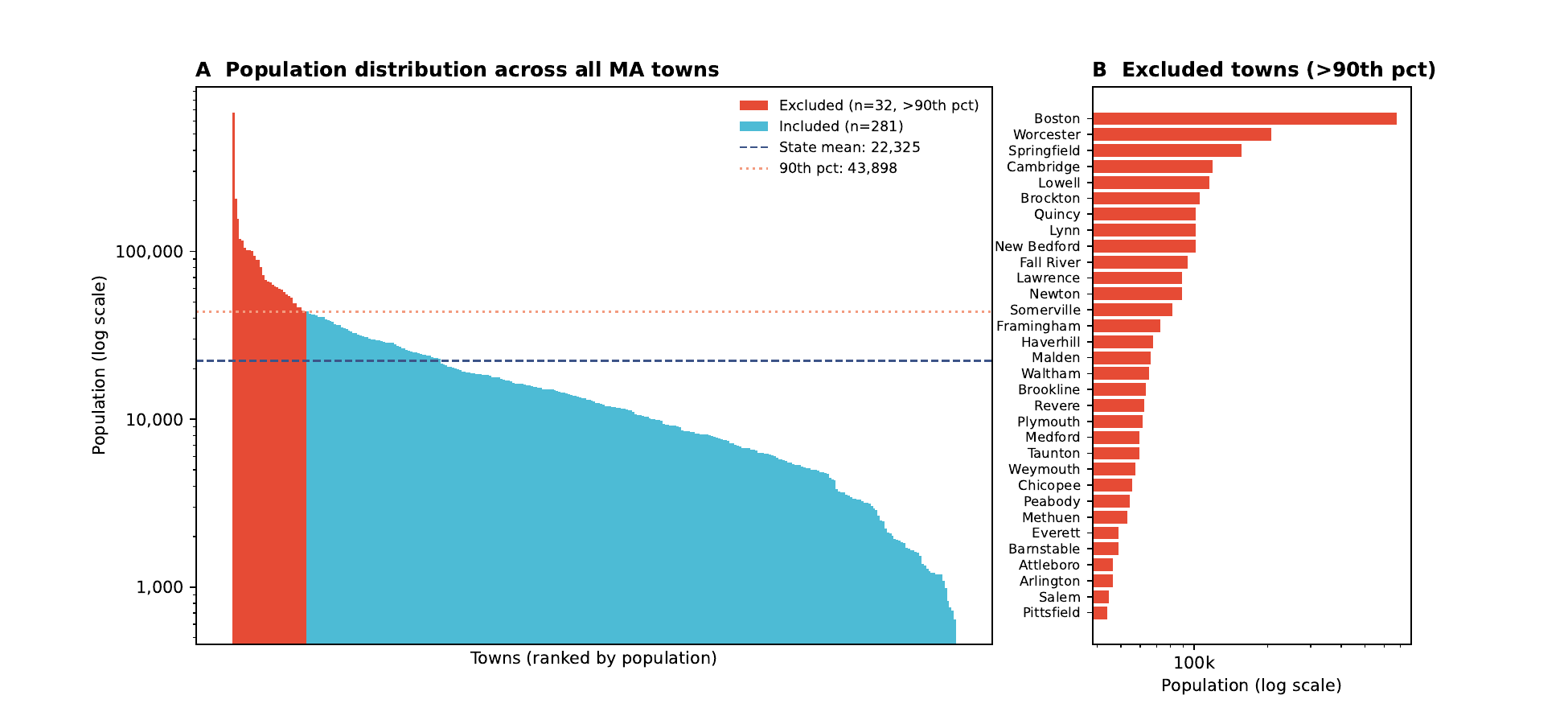}
\caption{Population distribution of Massachusetts towns and robustness sample construction. (A) Population of all 313 Massachusetts towns included in the primary analysis, ranked in descending order (log scale). Towns above the 90th population percentile are highlighted in red and excluded from the robustness analysis; the remaining towns are shown in teal. The navy dashed line indicates the unweighted state mean across all towns. (B) The 32 excluded high-population municipalities, ranked by population. Notable exclusions include Boston, Worcester, and Springfield, outliers whose scale and urban density may confound mobility responses relative to the broader Massachusetts town distribution. The remaining 281 towns constitute the estimation sample for the robustness check in Section~\ref{suppsubsec:robustness-exclude-towns}. Excluding these high-population outliers leaves the main results substantively unchanged (see Tables~\ref{tab:models_with_excluded_towns} and ~\ref{tab:models2_with_excluded_towns}).}
\label{fig:excluded_towns}
\end{figure}

We re-estimate the main models on a sample that drops the most populous municipalities, to confirm the spillover is not an artifact of a few large towns. Removing every Massachusetts municipality above the 90th population percentile takes out 32 of the 313 towns for the full study period, leaving 281 towns and 14{,}612 town-week observations. The excluded towns and their populations are shown in Fig.~\ref{fig:excluded_towns}, and the estimates are reported in Tables~\ref{tab:models_with_excluded_towns} and~\ref{tab:models2_with_excluded_towns}. They agree with the main results: the behavioral spillover is not driven by high-population municipalities.

% Robustness Check Table 1
% Dependent variable: non-home dwell time deviation from pre-COVID baseline (minutes/day)
% Sample: Massachusetts towns excluding those above the 90th percentile of population
%         (32 high-population municipalities removed, e.g.\ Boston, Worcester, Springfield)
%         Remaining sample: 281 towns, N = 14,612 town-week observations.
% SE clustered at the town level. All models include town and week fixed effects.

\begin{table}[htbp]
\centering
\caption{Robustness Check I: Excluding High-Population Municipalities. These models replicate the main finding of the paper, with one major change to the data, the removal of high-population municipalities. The dependent variable is $Y^{adj}_{it}$, the weekly median non-home dwell time for town $i$ in week $t$ expressed as a deviation from that town's pre-COVID baseline (minutes per day). The sample excludes the 32 Massachusetts municipalities above the 90th percentile of population (including Boston, Worcester, and Springfield), leaving 281 towns and 14{,}612 town-week observations. $Y^{adj}_{i,t-1}$ is the one-week lag of the dependent variable, capturing mobility persistence. $X_{it}^{\text{per10k}}$ is the weekly COVID-19 case count per 10{,}000 residents in town $i$, measuring local epidemic severity. $S^{\mathrm{PreG}}_{i,t-1}$ is the pre-COVID inter-town mobility network (origin-normalized) multiplied by the vector of lagged mobility deviations across all towns, yielding a network-weighted average of neighboring towns' lagged mobility, the key \emph{behavioral spillover} (``bubble formation'') term. All models include town and week two-way fixed effects. Standard errors clustered at the town level are in parentheses. $^{*}p<0.10$, $^{**}p<0.05$, $^{***}p<0.01$. }
\label{tab:models_with_excluded_towns}

\footnotesize % Standard for tables in high-impact journals
\setlength{\tabcolsep}{0pt} % Let LaTeX calculate the exact spacing
\begin{tabular*}{\textwidth}{@{\extracolsep{\fill}} l ccc ccc cc}
\toprule
 & \multicolumn{8}{c}{\textit{Dep. var.: Non-home dwell time deviation (min/day)}} \\
\cmidrule(lr){2-9}
 & (1) & (2) & (3) & (4) & (5) & (6) & (7) & (8) \\
\midrule
$Y^{adj}_{i,t-1}$ & & & & & $0.460^{***}$ & $0.459^{***}$ & $0.454^{***}$ & $0.453^{***}$ \\
 & & & & & $(0.048)$ & $(0.048)$ & $(0.049)$ & $(0.049)$ \\[3pt]

$X_{it}^{\text{per10k}}$ & & $-0.139^{***}$ & & $-0.126^{***}$ & & $-0.083^{***}$ & & $-0.077^{***}$ \\
 & & $(0.045)$ & & $(0.045)$ & & $(0.026)$ & & $(0.026)$ \\[3pt]

$S^{\mathrm{PreG}}_{i,t-1}$ & & & $0.531^{**}$ & $0.521^{**}$ & & & $0.250^{**}$ & $0.244^{**}$ \\
 & & & $(0.214)$ & $(0.212)$ & & & $(0.101)$ & $(0.101)$ \\[3pt]

Intercept & $-221.5^{***}$ & $-218.9^{***}$ & $-193.3^{***}$ & $-191.5^{***}$ & $-116.4^{***}$ & $-115.1^{***}$ & $-104.4^{***}$ & $-103.5^{***}$ \\
 & $(2.805)$ & $(3.407)$ & $(12.053)$ & $(12.023)$ & $(11.917)$ & $(12.093)$ & $(12.178)$ & $(12.282)$ \\
\midrule
$R^2$ & 0.756 & 0.756 & 0.758 & 0.759 & 0.810 & 0.810 & 0.811 & 0.811 \\
$N$ & 14,612 & 14,612 & 14,612 & 14,612 & 14,612 & 14,612 & 14,612 & 14,612 \\
Town FE & Yes & Yes & Yes & Yes & Yes & Yes & Yes & Yes \\
Week FE & Yes & Yes & Yes & Yes & Yes & Yes & Yes & Yes \\
\bottomrule
\addlinespace[2pt]
\multicolumn{9}{p{\textwidth}}{\scriptsize \textit{Notes:} Models 1--4 are static two-way FE specifications; Models 5--8 augment these with the lagged dependent variable. Standard errors (in parentheses) clustered at the town level. Significance: $^{***} p<0.01$, $^{**} p<0.05$, $^{*} p<0.10$.}
\end{tabular*}
\end{table}

\begin{table}[htbp]
\centering
\caption{Robustness Check II: Excluding High-Population Municipalities (Two-Way Clustered SE).
  The dependent variable is $Y^{adj}_{it}$, the weekly median non-home dwell time for town $i$ 
  in week $t$ expressed as a deviation from that town's pre-COVID baseline (minutes per day). 
  The sample excludes the 32 Massachusetts municipalities above the 90th percentile of 
  population (including Boston, Worcester, and Springfield), leaving 281 towns and 14,612 
  town-week observations. $Y^{adj}_{i,t-1}$ is the one-week lag of the dependent variable. 
  $X_{it}^{\text{per10k}}$ is the weekly COVID-19 case count per 10,000 residents in town $i$. 
  $S^{\mathrm{PreG}}_{i,t-1}$ is the pre-COVID network-weighted average of 
  neighboring towns' lagged mobility. $S^{\mathrm{PreG}\cdot X}_{i,t-1}$ 
  is the pre-COVID network-weighted average of neighboring towns' lagged COVID-19 case rates. 
  Models 1--4 include town and week two-way fixed effects; Models 5--8 omit fixed effects but 
  include $Y^{adj}_{i,t-1}$ to absorb serial dependence. Standard errors are two-way clustered 
  at the town and week level. $^{*}p<0.10$, $^{**}p<0.05$, $^{***}p<0.01$.}
\label{tab:models2_with_excluded_towns}

\footnotesize
\setlength{\tabcolsep}{0pt}
\begin{tabular*}{\textwidth}{@{\extracolsep{\fill}} l ccc ccc cc}
\toprule
 & \multicolumn{8}{c}{\textit{Dep. var.: Non-home dwell time deviation from pre-COVID baseline (min/day)}} \\
\cmidrule(lr){2-9}
 & (1) & (2) & (3) & (4) & (5) & (6) & (7) & (8) \\
\midrule
$Y^{adj}_{i,t-1}$ & & & & & $0.4598^{***}$ & $0.4586^{***}$ & $0.4542^{***}$ & $0.4532^{***}$ \\
 & & & & & $(0.0476)$ & $(0.0475)$ & $(0.0487)$ & $(0.0485)$ \\[4pt]
$X_{it}^{\text{per10k}}$ & & & $-0.1397^{***}$ & $-0.1259^{***}$ & & $-0.0827^{***}$ & & $-0.0773^{***}$ \\
 & & & $(0.0417)$ & $(0.0448)$ & & $(0.0262)$ & & $(0.0264)$ \\[4pt]
$S^{\mathrm{PreG}}_{i,t-1}$ & & $0.5312^{**}$ & & $0.5209^{**}$ & & & $0.2495^{**}$ & $0.2438^{**}$ \\
 & & $(0.2136)$ & & $(0.2120)$ & & & $(0.1009)$ & $(0.1006)$ \\[4pt]
$S^{\mathrm{PreG}\cdot X}_{i,t-1}$ & $-0.1045$ & & $0.0196$ & & & & & \\
 & $(0.2614)$ & & $(0.2302)$ & & & & & \\[4pt]
Intercept & $-221.03^{***}$ & $-193.29^{***}$ & $-218.94^{***}$ & $-191.45^{***}$ & $-116.36^{***}$ & $-115.07^{***}$ & $-104.39^{***}$ & $-103.46^{***}$ \\
 & $(3.4203)$ & $(12.0531)$ & $(3.8220)$ & $(12.0227)$ & $(11.9168)$ & $(12.0929)$ & $(12.1780)$ & $(12.2816)$ \\
\midrule
$R^2$ & 0.756 & 0.758 & 0.756 & 0.759 & 0.810 & 0.810 & 0.811 & 0.811 \\
$\bar{R}^2$ & 0.750 & 0.753 & 0.751 & 0.753 & 0.806 & 0.806 & 0.806 & 0.806 \\
$N$ & 14,612 & 14,612 & 14,612 & 14,612 & 14,612 & 14,612 & 14,612 & 14,612 \\
Town FE & Yes & Yes & Yes & Yes & No & No & No & No \\
Week FE & Yes & Yes & Yes & Yes & No & No & No & No \\
Clustered SE & T \& W & T \& W & T \& W & T \& W & T \& W & T \& W & T \& W & T \& W \\
\bottomrule
\addlinespace[2pt]
\multicolumn{9}{p{\textwidth}}{\scriptsize \textit{Notes:} This table replicates Table~\ref{tab:models_with_excluded_towns} using two-way clustering (town and week). It additionally introduces $S^{\mathrm{PreG}\cdot X}_{i,t-1}$, the pre-COVID network-weighted lagged COVID case rate of connected towns. The coefficient on $S^{\mathrm{PreG}\cdot X}_{i,t-1}$ is statistically indistinguishable from zero (Models 1 and 3), indicating that awareness of COVID incidence in socially connected municipalities does not independently suppress local mobility. By contrast, $S^{\mathrm{PreG}}_{i,t-1}$ remains positive and significant, supporting a social-learning or norm-conformity mechanism rather than a pure risk-information channel. Models 5--8 replace two-way FE with the lagged dependent variable. Standard errors in parentheses. $^{***}p<0.01$, $^{**}p<0.05$, $^{*}p<0.10$.}
\end{tabular*}
\end{table}

\section*{Demographic decomposition: robustness for the within-mobility-community result}\label{suppsec:demographic-robustness}

This section reports the demographic-decomposition analyses that supplement the within-mobility-community spillover finding reported in the main text. They serve as a robustness test: if the spillover identified along the mobility partition were spurious, or carried merely by some unmodelled correlate of mobility-community membership, then a decomposition along a separately constructed demographic axis should disrupt the within / across contrast. It does not. Instead, the spillover is sharpest where the two axes intersect, among town pairs that both belong to the same mobility community and share a demographic profile, and it is absorbed by a continuous demographic-similarity modulator built on the matrix $D$ defined in §\ref{suppsubsec:methods-D}.

\subsection*{Demographic-Louvain partition and joint structure with mobility}

The demographic-community partition (main-text Fig.~\ref{fig:partitions}c) recovers eight communities at $Q = 0.86$. Within-demographic $\mathrm{PreG}$ edges are only $1.4\times$ heavier than across-demographic edges ($n_w = 4{,}152$, $n_b = 22{,}508$ pairs; $d_{\log_{10}} = 0.19$; main-text Fig.~\ref{fig:partitions}d): demographic similarity is not a proxy for connectivity. The corresponding within / across decomposition of $S^{\mathrm{PreG}}_{i,t-1}$ along the demographic partition (main-text Table~\ref{tab:results_main}, columns~5--6) recovers a different pattern from the mobility partition: both the within- and across-demographic spillovers enter positively and significantly, in sharp contrast to the clean within-only signature obtained under the mobility partition. Auxiliary specifications that re-introduce the unsplit network spillover (Table~\ref{tab:si_demographic_partition}) sharpen the contrast: under the demographic partition the across term collapses while the unsplit $S^{\mathrm{PreG}}_{i,t-1}$ remains significant even with persistence absorbed, whereas under the mobility partition the unsplit term is absorbed by the within mask. The demographic partition tracks behavior partially but does not, on its own, carve up the network channel.

\subsection*{Joint mobility-by-demographic decomposition}\label{suppsubsec:four-mask}

The two partitions are essentially \emph{orthogonal} (main-text Fig.~\ref{fig:partitions}f): on the 313-town intersection, $\mathrm{ARI} = 0.082$ and $\mathrm{NMI} = 0.144$. Mobility communities and demographic communities pick up distinct dimensions of inter-town structure, so neither partition is a coarsening of the other. We therefore decompose $\mathrm{PreG}$ into four Hadamard masks indexed by whether each unit pair is (within / across) the mobility partition and (within / across) the demographic partition: $\mathrm{PreG}^{mm}, \mathrm{PreG}^{ma}, \mathrm{PreG}^{am}, \mathrm{PreG}^{aa}$. Each probes a structurally distinct kind of peer pair. The mass of social influence loads on the within--within block (Table~\ref{tab:results_double_partition}): the $\mathrm{PreG}^{mm}$ coefficient is roughly three times the headline pooled estimate, the within-mobility / across-demographic block is positive but smaller, and the across-mobility blocks are statistically indistinguishable from zero throughout. The spillover concentrates among pairs that share both mobility-community membership and demographic similarity; behavioral diffusion requires the joint condition.

\begin{table}[!htbp]
\centering
\caption{\textbf{Joint mobility-by-demographic decomposition (four Hadamard masks).} $\mathrm{PreG}$ is split by crossing the mobility-Louvain partition (within $w$ / across $a$) with the demographic-Louvain partition: $\mathrm{PreG}^{mm}$ is within-mobility \emph{and} within-demographic, $\mathrm{PreG}^{ma}$ is within-mobility, across-demographic, $\mathrm{PreG}^{am}$ is across-mobility, within-demographic, and $\mathrm{PreG}^{aa}$ is across-mobility, across-demographic. All columns include town and week fixed effects; standard errors clustered at the town level. $^{*}p<0.10$, $^{**}p<0.05$, $^{***}p<0.01$.}
\label{tab:results_double_partition}
\footnotesize
\setlength{\tabcolsep}{0pt}
\begin{tabular*}{\linewidth}{@{\extracolsep{\fill}} l cccc}
\toprule
 & \multicolumn{4}{c}{\textit{Dep.\ var.: $Y^{adj}_{i,t}$ (min/day)}} \\
\cmidrule(lr){2-5}
 & (1) & (2) & (3) & (4) \\
 & FE, 4 masks & FE, 4 masks + $S^{\mathrm{PreG}}$ & FE+AR, 4 masks & FE+AR, 4 masks + $S^{\mathrm{PreG}}$ \\
\midrule
$Y^{adj}_{i,t-1}$ & & & $0.455^{***}$ & $0.455^{***}$ \\
 & & & $(0.047)$ & $(0.047)$ \\[3pt]

$X_{i,t}$ (per 10k) & $-0.108^{***}$ & $-0.108^{***}$ & $-0.068^{***}$ & $-0.068^{***}$ \\
 & $(0.042)$ & $(0.042)$ & $(0.025)$ & $(0.025)$ \\[3pt]

$S^{\mathrm{PreG}}_{i,t-1}$ & & $0.414^{*}$ & & $0.075$ \\
 & & $(0.232)$ & & $(0.124)$ \\[3pt]

$\mathrm{PreG}^{mm} \cdot Y^{adj}_{t-1}$ & $1.542^{***}$ & $1.128^{**}$ & $0.640^{***}$ & $0.564^{***}$ \\
 & $(0.468)$ & $(0.442)$ & $(0.228)$ & $(0.217)$ \\[3pt]

$\mathrm{PreG}^{ma} \cdot Y^{adj}_{t-1}$ & $0.412^{**}$ & $-0.002$ & $0.191^{**}$ & $0.115$ \\
 & $(0.165)$ & $(0.252)$ & $(0.078)$ & $(0.130)$ \\[3pt]

$\mathrm{PreG}^{am} \cdot Y^{adj}_{t-1}$ & $0.248$ & $-0.166$ & $-0.448$ & $-0.523$ \\
 & $(1.156)$ & $(0.988)$ & $(0.623)$ & $(0.530)$ \\[3pt]

$\mathrm{PreG}^{aa} \cdot Y^{adj}_{t-1}$ & $-0.132$ & $-0.546$ & $-0.006$ & $-0.081$ \\
 & $(0.410)$ & $(0.437)$ & $(0.218)$ & $(0.240)$ \\
\midrule
$R^{2}$            & 0.773 & 0.773 & 0.822 & 0.822 \\
$N$ (town-weeks)   & 16{,}276 & 16{,}276 & 16{,}276 & 16{,}276 \\
Town FE            & Yes & Yes & Yes & Yes \\
Week FE            & Yes & Yes & Yes & Yes \\
\bottomrule
\end{tabular*}
\vspace{4pt}
{\scriptsize \textit{Notes.} The within-mobility, within-demographic block ($\mathrm{PreG}^{mm}$) carries the spillover across all four specifications and is the only block that survives both the addition of the unsplit network spillover (columns~2,\,4) and the lagged outcome (columns~3,\,4). The across-mobility blocks ($\mathrm{PreG}^{am}$, $\mathrm{PreG}^{aa}$) are statistically indistinguishable from zero throughout: behavioral diffusion does not bridge mobility communities even when the connected towns share demographics. Consistent with the main-text within-mobility-community result, the channel is concentrated where the two structures intersect.}
\end{table}

\subsection*{Continuous demographic-similarity modulator and horse-race}\label{suppsubsec:horse-race}

The four-mask decomposition relies on a binary demographic partition; a more rigorous robustness check is the continuous $313\times 313$ similarity matrix $D$ defined in §\ref{suppsubsec:methods-D}. Replacing the binary mask with $D$ collapses the demographic structure into a single weight on each town pair. We then construct the modulated spillover regressor of Eq.~\eqref{eq:s_pregxd_SI}, which retains every mobility-network edge but reweights it by the demographic similarity of the two endpoints. A high-flow pair of demographically distant towns now contributes less than a high-flow pair of demographically similar towns.

The horse-race in Table~\ref{tab:results_demographic} enters the plain mobility signal $S^{\mathrm{PreG}}$ against the modulated form $S^{\mathrm{PreG}\odot D}$ in the same regression. Entered alone, both spillovers are positive and significant (columns~1--2). Once the modulated form is added to a specification that already contains the plain spillover, the plain signal collapses to insignificance while the modulated term enters strongly (column~3); the same pattern holds with $S^{D}$ added (column~4) and with the lagged outcome absorbed (column~5).

\begin{table}[!htbp]
\centering
\caption{\textbf{Demographic similarity modulates the mobility-network spillover (continuous-modulator robustness).} Outcome and fixed-effects structure as in main-text Table~\ref{tab:results_main}. $D$ is the $313\times313$ town-by-town demographic-similarity matrix from §\ref{suppsubsec:methods-D}, mapped to similarity through an RBF kernel. $S^{D}_{i,t-1}$ is the demographic-similarity-weighted lagged behavior, and $S^{\mathrm{PreG}\odot D}_{i,t-1}$ is the multiplicative spillover defined in Eq.~\eqref{eq:s_pregxd_SI}. Once the modulated term enters, the plain mobility spillover loses significance and stays insignificant after adding $S^{D}$ and the lagged outcome. $^{*}p<0.10$, $^{**}p<0.05$, $^{***}p<0.01$.}
\label{tab:results_demographic}
\footnotesize
\setlength{\tabcolsep}{0pt}
\begin{tabular*}{\linewidth}{@{\extracolsep{\fill}} l ccccc}
\toprule
 & \multicolumn{5}{c}{\textit{Dep.\ var.: $Y^{adj}_{i,t}$ (min/day)}} \\
\cmidrule(lr){2-6}
 & (1) & (2) & (3) & (4) & (5) \\
 & PreG & D only & PreG vs $\mathrm{PreG}\odot D$ & PreG + D + $\mathrm{PreG}\odot D$ & AR(1) horse-race \\
\midrule
$Y^{adj}_{i,t-1}$ & & & & & $0.452^{***}$ \\
 & & & & & $(0.047)$ \\[3pt]

$X_{i,t}$ (per 10k) & $-0.110^{***}$ & $-0.128^{***}$ & $-0.110^{***}$ & $-0.110^{***}$ & $-0.069^{***}$ \\
 & $(0.042)$ & $(0.043)$ & $(0.041)$ & $(0.042)$ & $(0.024)$ \\[3pt]

$S^{\mathrm{PreG}}_{i,t-1}$ & $0.512^{***}$ & & $0.117$ & $0.122$ & $0.080$ \\
 & $(0.191)$ & & $(0.115)$ & $(0.109)$ & $(0.064)$ \\[3pt]

$S^{D}_{i,t-1}$ & & $0.0017^{***}$ & & $0.0001$ & \\
 & & $(0.0006)$ & & $(0.0007)$ & \\[3pt]

$S^{\mathrm{PreG}\odot D}_{i,t-1}$ & & & $1.594^{***}$ & $1.569^{***}$ & $0.608^{***}$ \\
 & & & $(0.364)$ & $(0.415)$ & $(0.173)$ \\
\midrule
$R^{2}$            & 0.772 & 0.771 & 0.775 & 0.775 & 0.822 \\
$N$ (town-weeks)   & 16{,}276 & 16{,}276 & 16{,}276 & 16{,}276 & 16{,}276 \\
Town FE            & Yes & Yes & Yes & Yes & Yes \\
Week FE            & Yes & Yes & Yes & Yes & Yes \\
\bottomrule
\end{tabular*}
\vspace{4pt}
{\scriptsize \textit{Notes.} Column 1 reproduces the baseline mobility spillover from main-text Table~\ref{tab:results_main}. Column 2 reports a purely demographic spillover $S^{D}$. Column 3 is the headline horse-race: when $S^{\mathrm{PreG}\odot D}$ enters alongside $S^{\mathrm{PreG}}$, the plain mobility spillover collapses and is no longer significant, while the modulated channel enters strongly. Column 4 adds $S^{D}$ and the same pattern holds. Column 5 adds the lagged outcome and the modulated channel still survives.}
\end{table}

\subsubsection*{Collinearity caveat.}
Because $S^{\mathrm{PreG}}$ and $S^{\mathrm{PreG}\odot D}$ are built from the same network and the same lagged outcome vector, with the modulator $D$ only reweighting each edge by the demographic similarity of its endpoints, the two regressors are mechanically correlated. The collapse of $S^{\mathrm{PreG}}$ in column~3 could in principle reflect variance moving between two near-collinear terms rather than evidence that the modulated form is the operative channel. Two facts in the table argue against that reading. First, the pattern is asymmetric: when both regressors enter, the modulated form keeps its full standalone magnitude and significance while the plain form collapses, rather than both terms weakening as pure collinearity would predict. Second, the fit improves, with $R^{2}$ rising from $0.772$ (column~1, plain mobility alone) to $0.775$ (column~3, plain plus modulated) and holding at $0.775$ in column~4 after the demographic-only term is added. The gain is small in absolute terms but consistent with the modulated form capturing variance the plain form does not. 

\subsubsection*{Implication of the demographic robustness.}
Read together, the four-mask decomposition and the continuous-modulator horse-race deliver the same conclusion: behavioral spillover along the mobility network concentrates among town pairs that are also demographically similar. The within-mobility-community result in the main text is therefore not an artifact of the mobility partition: it is reinforced by every demographic-decomposition variant we have run. Read structurally, the result says that interaction structure (where towns connect) and similarity structure (which towns resemble each other) are both relevant for behavioral diffusion, and that the joint condition concentrates the channel more sharply than either alone.

\section*{Supplementary Discussion}
The supplementary analyses reinforce the three claims that carry the main text. The greedy-modularity baseline recovers the same community structure as Louvain, so the partition is not an artifact of one algorithm. Excluding the high-population municipalities leaves the spillover unchanged, so it is not an artifact of Boston or the other large towns. And the demographic decompositions, both the four-mask crossing and the continuous-modulator horse-race, place the channel where mobility and similarity overlap rather than along either axis alone. Together they support reading the residual cross-community variation in behavior as a structured, community-level quantity rather than noise.

\end{document}